\documentclass[article]{revtex4-1}
\usepackage{graphicx}
\usepackage{booktabs}
\usepackage{float}
\usepackage{bm}
\usepackage{graphicx}
\usepackage{gensymb}
\usepackage{csvsimple}
\usepackage[figuresright]{rotating}
\usepackage{multirow}
\usepackage[export]{adjustbox}
\usepackage{subfig}
\usepackage{url}
\begin{document}
\preprint{PRL/123-SM}
\title{Understanding the chemical coupling between ground-level ozone and oxides of nitrogen in ambient air at industrial and commercial sites in central India}
\author{Suchetana Sadhukhan}
\email{s.suchetana@gmail.com}
\affiliation{Department of Physics, School of Advanced Sciences and Languages, VIT Bhopal University, Bhopal, Madhya Pradesh - 466114, India.}
\author{Satish Bhagwatrao Aher}%
\email{aher.satish@icmr.gov.in}
\affiliation{Division of Environmental Monitoring and Exposure Assessment (Air), ICMR-National
Institute for Research in Environmental Health, Bhopal, Madhya Pradesh - 462030, India.}
\date{\today}
\author{Pon Harshavardhanan}
\email{pon.harshavardhanan@vitbhopal.ac.in }
\affiliation{School of Computing Science \& Engineering, VIT Bhopal University, Bhopal, Madhya Pradesh - 466114, India.}
%\equalcont{These authors contributed equally to this work.}
\author{Dharma Raj}
\email{discoverability.drp@gmail.com}
\affiliation{Division of Biostatistics and Bioinformatics, ICMR-National
Institute for Research in Environmental Health, Bhopal, Madhya Pradesh - 462030, India.
}
%\equalcont{These authors contributed equally to this work.}
\author{Subroto Shambhu Nandi}
\email{s.nandi76@icmr.gov.in}
\affiliation{Division of Environmental Monitoring and Exposure Assessment (Air), ICMR-National
Institute for Research in Environmental Health, Bhopal, Madhya Pradesh - 462030, India.
}
%\onecolumn

%%==================================%%
%% sample for unstructured abstract %%
%%==================================%%
\begin{abstract}

This study investigates the hourly concentrations of Ground-level Ozone $(O_3)$, Nitric Oxide ($NO$), Nitrogen Dioxide ($NO_2$), and Oxides of Nitrogen ($NO_x$) in ambient air, along with the various meteorological parameters viz., ambient temperature, relative humidity, wind speed, and solar radiation over one year, from August $2020$ to July $2021$ for an industrial and commercial site in Madhya Pradesh, a central state in India. 
We also analyze the chemical coupling between the $O_3$ and ambient $NO$, $NO_2$, and $NO_x$ at both sites during the daytime. We focus on understanding how the concentration of the oxidant $OX$ (a combination of ozone and nitrogen dioxide) changes in relation to levels of $NO_x$ (a combination of nitrogen oxides) to determine whether the atmospheric sources of $OX$ are dependent on the  $NO_x$-independent or regional contributions and $NO_x$-dependent or local contributions. We also observe a significant positive correlation of $O_3$ with ambient temperature  and solar radiation but a strong negative correlation with relative humidity for the considered period.
We found the monthly variations of the pollutants' concentrations show a strong seasonality dependence. $O_3$ concentration becomes highest/lowest during summer/monsoon for both sites. In contrast, $NO_x$ exhibits the maximum and minimum concentration during monsoon and summer for both sites. The daily variation shows the opposite trend for $O_3$ and $NO_x$. $O_3$ reaches a peak at mid-day around $14:00$ Hrs when ambient temperature also goes to maximum and has the minimum value at night time due to lack of sunlight. 
\end{abstract}


\maketitle
\section{Introduction}\label{sec1}
For the last few years, a rapid increase in Ground-level Ozone ($O_3$) concentration throughout the world has gained the attention of researchers and Governments worldwide \cite{vin04, paoletti14, zhang02, geddes09}) in making effective policies to reduce $O_3$ levels in the future. India, a large and populous country, with a population of over 1.2 billion people according to the 2011 Census, is not also an exception. The main reason behind this scenario is fast industrialization and urbanization \cite{dhanya2021, shukla2021, notario12}. Until few years ago, mainly the Indo-Gangetic plain was known to be ``more polluted" \cite{mahapatra2014} which now has spread to other states as well, such as Maharashtra and Madhya Pradesh\footnote{Gandhiok, Jasjeev,  ``Average Madhya Pradesh resident losing $2.9$ years of life due", The Times of India, $02$nd September, 2021; \protect\url{http://timesofindia.indiatimes.com/articleshow/85847190.cms?utm_source=contentofinterest&utm_medium=text&utm_campaign=cppst}}. According to the Air Quality Life Index (AQLI) by the Energy Policy Institute of the University of Chicago, high levels of air pollution in Madhya Pradesh, a state in India, have been associated with a reduction in average life expectancy by 2.5 to 2.9 years. This highlights the importance of addressing air pollution and taking steps to improve air quality in the region\footnote{ \protect\url{https://aqli.epic.uchicago.edu/}}. 
Several researchers have already pointed out the various health risks associated to exposure with $O_3$, among which respiratory disease is the most significant one to mention  \cite{jerrett09, kim2020, rajak2020short, dhanya2021, shukla2021, singh2016, jayaraman2007, ware2016, conibear2018, pandey2021}. Not only to human, high level of $O_3$ also affects plants which can lead to extreme economic losses in the near future \cite{oksanen13, van09}.
Chemical association between $O_3$ and Oxides of nitrogen $(NO_x)$ is also important for the formation of $O_3$ and has been studied by many researchers \cite{yang05, mazzeo05, itano07, pudasainee06, gasmi2017, tiwari2015}. 
Oxides of nitrogen refer to the mixture of all the gases that are composed of nitrogen and oxygen, most significantly, nitric oxide $(NO)$ and nitrogen dioxide $(NO_2)$, also nitrous oxide $(N_2O)$, nitrogen pentoxide $(N_2O_5)$, and volatile organic compounds (VOCs) in a lesser ratio. $NO_x$ and VOCs are two common precursors for the formation of $O_3$ in the Earth's atmosphere. The non-linear dependency of $O_3$ on its precursors makes it the central importance of the investigation to give an effective strategy to reduce the $O_3$ level. In this regard, studies have already identified two regimes for the formation of $O_3$: dependent and independent of $NO_x$ concentration (or dependent on VOCs) \cite{sillman99}. It is also well known that $O_3$ formation rate depends on the amount of sunlight and ambient temperature (AT), hence it reaches to a peak concentration during mid-day \cite{londhe08}. Conversely, the concentration of $O_3$ decreases with high relative humidity \cite{CAMALIER20077127, li2021large}. However, a comparative assessment and magnitude of changes at diverse sources is lacking. A similar study has been reported by Nishant et al. \cite{nishanth12} for the southern part of India (Kannur, Kerala). However, present study is unique because it is the first to examine these interactions in the central region of India, which has a distinct meteorological profile and emission profile compared to other parts of the country. The study focuses on Madhya Pradesh, a central state in India, because it has relatively stable weather conditions, which can help to produce more reliable and conclusive results. By understanding the factors that influence ozone formation in this region, the study can inform policy decisions and help reduce ozone concentrations to protect public health.

\section{Data Description}\label{sec2}
\begin{figure}[H]
\centering
\includegraphics[width=\textwidth]{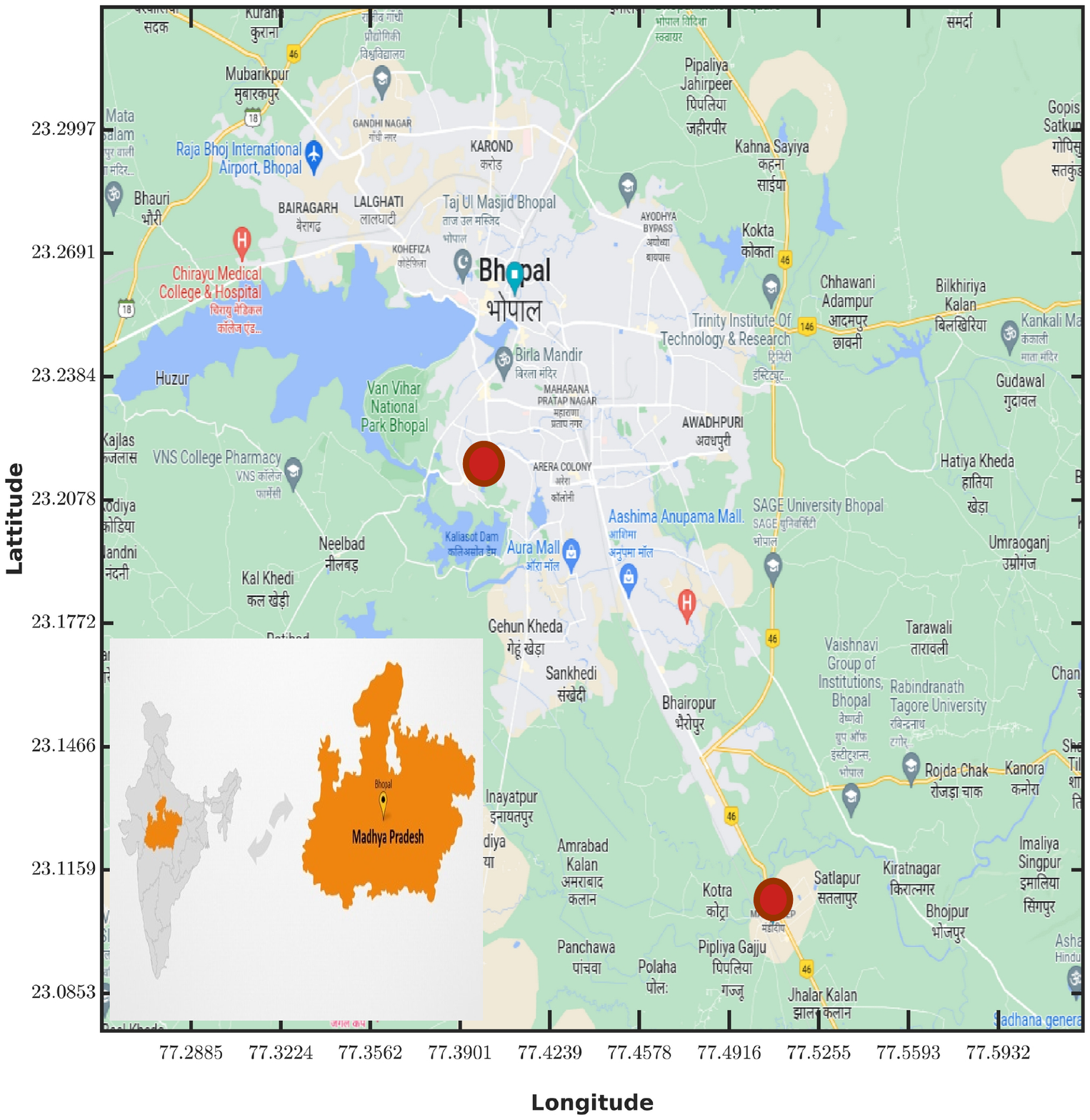}
\vspace{-6cm}
\caption{Map shows the location of the study areas (Bhopal and Mandideep). The red circles are the exact position of the pollutant measuring stations.}\label{fig1}
\end{figure}

For the analysis, real-time air pollution hourly data is considered from August $2020$ to July $2021$ for a commercial (Bhopal) and industrial (Mandideep) sites in Madhya Pradesh, a central state of India, located at an elevation of $90$ to $1,200$ meters above sea level. The air quality data, the hourly concentration of Nitrogen Dioxide ($NO_2$), Nitric oxide ($NO$), Ground level Ozone ($O_3$) and Oxides of Nitrogen ($NO_x$) are downloaded from the continuous ambient air quality monitoring stations operated by CPCB (Central Pollution Control Board)\footnote{\protect\url{https://app.cpcbccr.com/ccr//caaqm-dashboard-all/caaqm-landing}}. 
The red circle in Figure~\ref{fig1} shows the location of the pollutant measuring stations, and all the other geographical details are provided in Table~\ref{tab1}.
\begin{table}[h]
\begin{center}
\caption{Details of the continuous ambient air quality monitoring stations (CAAQMS) located at Bhopal and Mandideep.}\label{tab1}%
\begin{tabular}{@{}llll@{}}
\toprule
Station & Bhopal & Mandideep \\
\midrule
Area     & 463 			$km^2$             & 12.78 $km^2$  \\
Population Density    & 3900/$km^2$             & 4668/$km^2$     \\
Category   & Commercial             & Industrial   \\
Location  & 23.265° N, 77.412° E & 23.104° N, 77.511° E\\
\botrule
\end{tabular}
\footnotetext{Source(Bhopal): District Census Handbook – Bhopal" (PDF). Census of India. p.35. Archived (PDF) from the original on 7 August 2015. Retrieved 22 September 2015.}
\footnotetext{Source(Mandideep): Office of the Registrar General and Census Commissioner (web), Delimitation Commission of India (web), Rand McNally International Atlas 1994, School of Planning \& Architecture (web).}
\end{center}
\end{table}

At the central part of India, maximum temperature in summer reaches approximately $30-35\degree$C with less
humidity. Minimum temperatures during winter go down to around
 $15-20\degree$C with frequent fogs and low
visibility\footnote{\protect\url{https://en.climate-data.org/asia/india/madhya-pradesh/bhopal-2833/}}. Mandideep is a diverse and vibrant industrial town with a range of industries, e.g., manufacturing, chemical and petrochemical, food and beverage, textile and engineering, etc., with a considerable contribution to pollution due to industrial emissions. On the other hand, Bhopal is a center to a variety of commercial businesses and activities, including retail, finance, hospitality, real estate, agricultural activities in and around the city, emissions from transportation, construction of new buildings and infrastructure which leads to pollution.
\section{Methods}\label{secm}
For the analysis, we consider hourly data-set for one year from August $2020$ to July $2021$ of $O_3$, $NO$, $NO_2$, and $NO_x$. Data-set is refined first by removing the outliers. We consider the values greater than three standard deviation of average value over a window length of 24 hours, as outliers \cite{longmore2019comparison}. For our data-sets, we have some data-points missing. We replace them by taking the average of six data points (three before and three after) at the same hour of the day keeping the missing data point of any specific time at the center \cite{zainuri2015comparison}. This approach can help to improve the accuracy and reliability of the analysis by reducing the influence of outliers and filling in missing data. The meteorological conditions for Bhopal (commercial) and Mandideep (industrial) can be classified into main three seasons; Summer (March-June), monsoon (July-October) and winter (November-February). The concentrations of pollutants were calculated depending on the sunrise and sunset values according to the season, with the daytime period generally occurring between $6:00$ and $19:00$ Hrs in the summer, between $6:00$ and $18:00$ Hrs in the monsoon season, and between $7:00$ and $18:00$ Hrs in the winter.
 Dividing the data into seasons and daytime and nighttime concentrations by using the sunrise and sunset values for each season helps to understand how the concentrations of these pollutants vary across different seasons and meteorological conditions.

$O_3$ is a highly reactive gas formed through a chain of chemical reactions involving $NO_x$ and oxygen and influenced by a range of factors, including meteorological conditions, emissions, and Ultraviolet (UV) light. Hence, understanding the mechanisms of $O_3$ formation is necessary to address the impacts of air pollution on public health and the environment. The formation of $O_3$ is generally a dynamic process that involves the inter-conversions of $NO$, $NO_2$, $NO_x$ and, $O_3$ through photostationary state (PSS) reactions \cite{Leighton61, leighton2012}. $NO_2$ produces $NO$ and atomic oxygen via the photolysis process, and then, atomic oxygen reacts with oxygen to produce $O_3$, as mentioned in Reaction-$1$ and Reaction-$2$. $O_3$, as the product from Reaction-$2$, reacts with $NO$ and produces $NO_2$ (Reaction-$3$).
\begin{eqnarray*}
NO_2+h\nu \rightarrow NO+O \;\;\; \;\;\; \;\;\; \text{Reaction-1}\\
O+O_2+M \rightarrow O_3+M \;\;\; \;\;\; \;\;\; \text{Reaction-2}\\
NO+O_3 \rightarrow NO_2+O \;\;\; \;\;\; \;\;\; \text{Reaction-3}\\
\end{eqnarray*}
The three equations mentioned above are referred to as the null cycle
due to the very fast process leading to a steady-state cycle.   
Thus, the concentrations of $O_3$ during the PSS are given by the equation \cite{Leighton61}:
\begin{equation}
\frac{[NO][O_3]}{[NO_2]}=\frac{j_1}{k_3}
\end{equation}
The photolysis rate $(j_1)$ is the rate at which nitrogen dioxide $(NO_2)$ is dissociated into individual atoms of nitrogen and oxygen through the absorption of sunlight. $j_1$ depends on the solar radiation intensity. The rate coefficient $(k_3)$ measures the rate at which $NO$ reacts with $O_3$ to form $NO_2$ and $O_2$. The rate coefficient $k_3$ for Reaction-3 is directly proportional to the AT and can be calculated using the
following equation \cite{seinfeld98}:
\begin{equation}
k_3(ppm^{-1} min^{-1})= 3.23 \times 10^3 \times exp[-1430/T]
\label{eqn10}
\end{equation}
Here $T$ refers to the ambient temperature (AT).

To see the interrelations between the pollutants and various meteorological parameters and analyze the occurrence of
various pollutants we calculate the Pearson Correlation coefficient between the ambient pollutants \cite{latini2002influence}. 
\begin{equation}
%\rho(x,y)=\frac{\langle(x_i-\langle x \rangle) (y_i-\langle y \rangle)\rangle}{\sigma_x \sigma_y}
\rho(x,y)=\frac{cov(x,y)}{\sigma_x \sigma_y}
\label{eqn11}
\end{equation}
here, $x$ and $y$ are any two variables (e.g., pollutants’
concentration, meteorological parameter data) time-series and $cov$ stands for covariance. $\rho(x,y)$ derives the linear correlation between two
variables of $x$ and $y$, where it can take a minimum value as $-1$ showing a complete negative correlation, and $+1$ is a complete positive linear correlation \cite{lee1988thirteen}.

To analyze the chemical coupling between $O_3$, $NO$, and $NO_2$, we use the daylight concentrations of oxidant levels $(OX=O_3+NO_2)$ and $NO_x$, fitted to a linear regression $(mx + c)$ model. The slope of the line represents the rate of change of the $OX$ concentration with respect to the $NO_x$ concentration, and it can be thought of as the $NO_x$-dependent contribution to the $OX$ concentration. The y-intercept of the line represents the $OX$ concentration when the $NO_x$ concentration is zero, and it can be thought of as the $NO_x$-independent contribution to the $OX$ concentration \cite{clapp01, mazzeo05, kley1994}. The use of a linear regression model allows you to quantify the relationship between $OX$ and $NO_x$ concentrations and to understand how changes in one variable may affect the other. It can also provide insight into the mechanisms underlying the chemical coupling between $O_3$, $NO$, and $NO_2$. For $NO_x$-dependent contribution, the concentration of total oxidants depends on the local pollution due to variation in $NO_x$ concentration. On the other hand, $NO_x$-independent contribution represents the $OX$ concentration that is not directly related to $NO_x$ concentration, and it may be influenced by regional or background factors. Local contribution depends on the factors, including the prevalent local photochemistry, thus positively correlates with the concentration of $O_3$, $OX$, and, also, on the local sources, thus with $NO_x$, which strongly depends on the local emissions. 
\section{Results and Discussions}\label{sec3}
\subsection{Seasonal dependency of $O_3$, $NO$, $NO_2$, and $NO_x$}\label{sec31}
%\caption{Variation of daylight values of $[OX]$ with $NO_x$ for (a) Commercial and (b) Industrial.}\label{fig6} 
During summer, chemical production of $O_3$ gets increased for the availability of ideal meteorological conditions.  In winter, on the other hand, atmospheric stability is attained because of frequent inversions which helps accumulate pollutants near the surface (known as photochemical smog) \cite{tiwari14}. 
Due to the increase in the number of industries which leads to growing population activities and more vehicles on the road which results in higher concentration levels of $NO_x$ and VOCs, and increased ozone conversion rate obvious for Industrial areas. This also makes it clearly understandable the reason behind the lower $O_3$ concentration in Industrial compared to Commercial area. 
\begin{figure}[H]
\subfloat[]{\label{fig3:a}\includegraphics[width=0.45\textwidth]{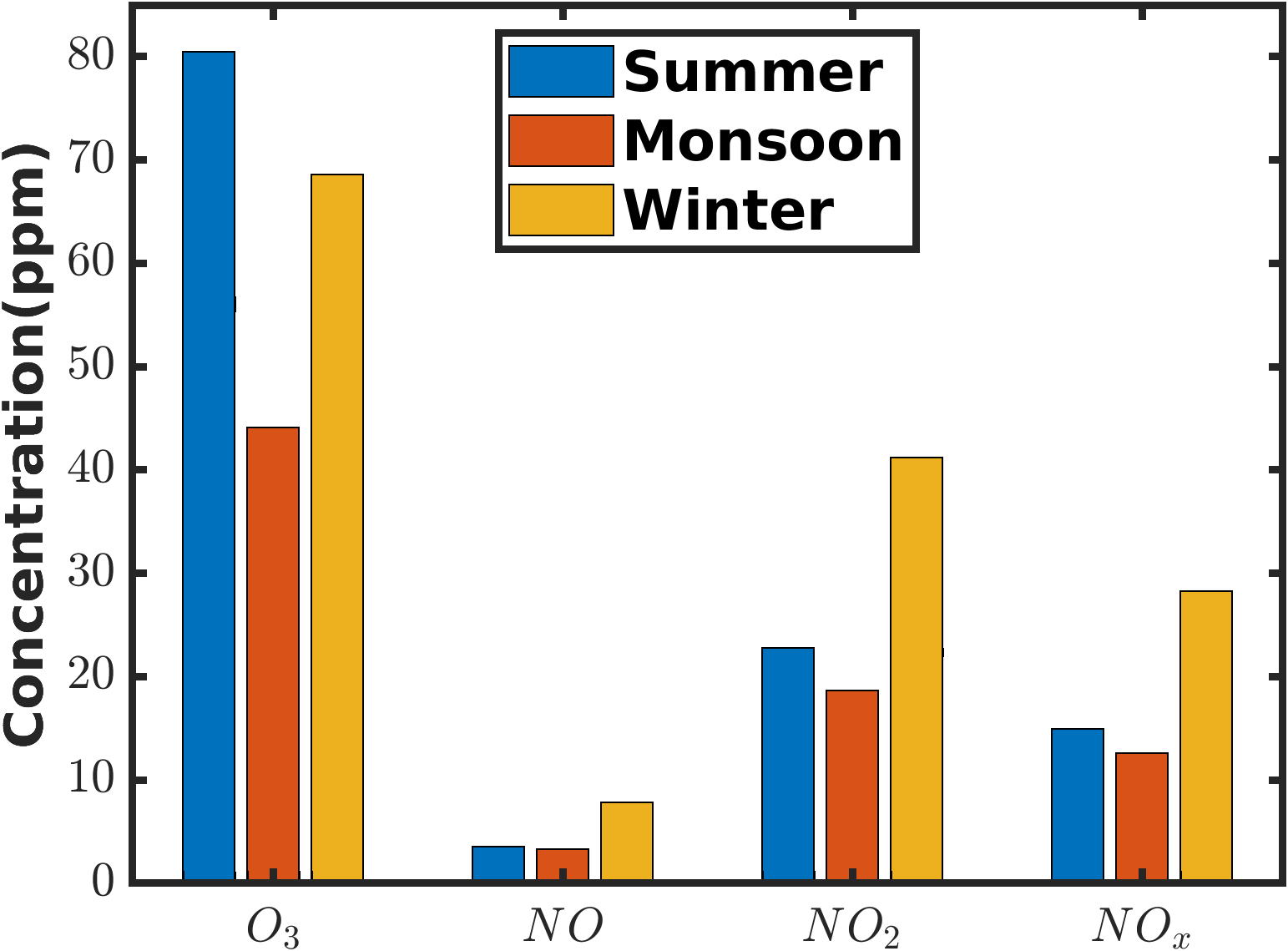}}%
\hspace{1cm}
  \subfloat[]{\label{fig3:b}\includegraphics[width=0.45\textwidth]{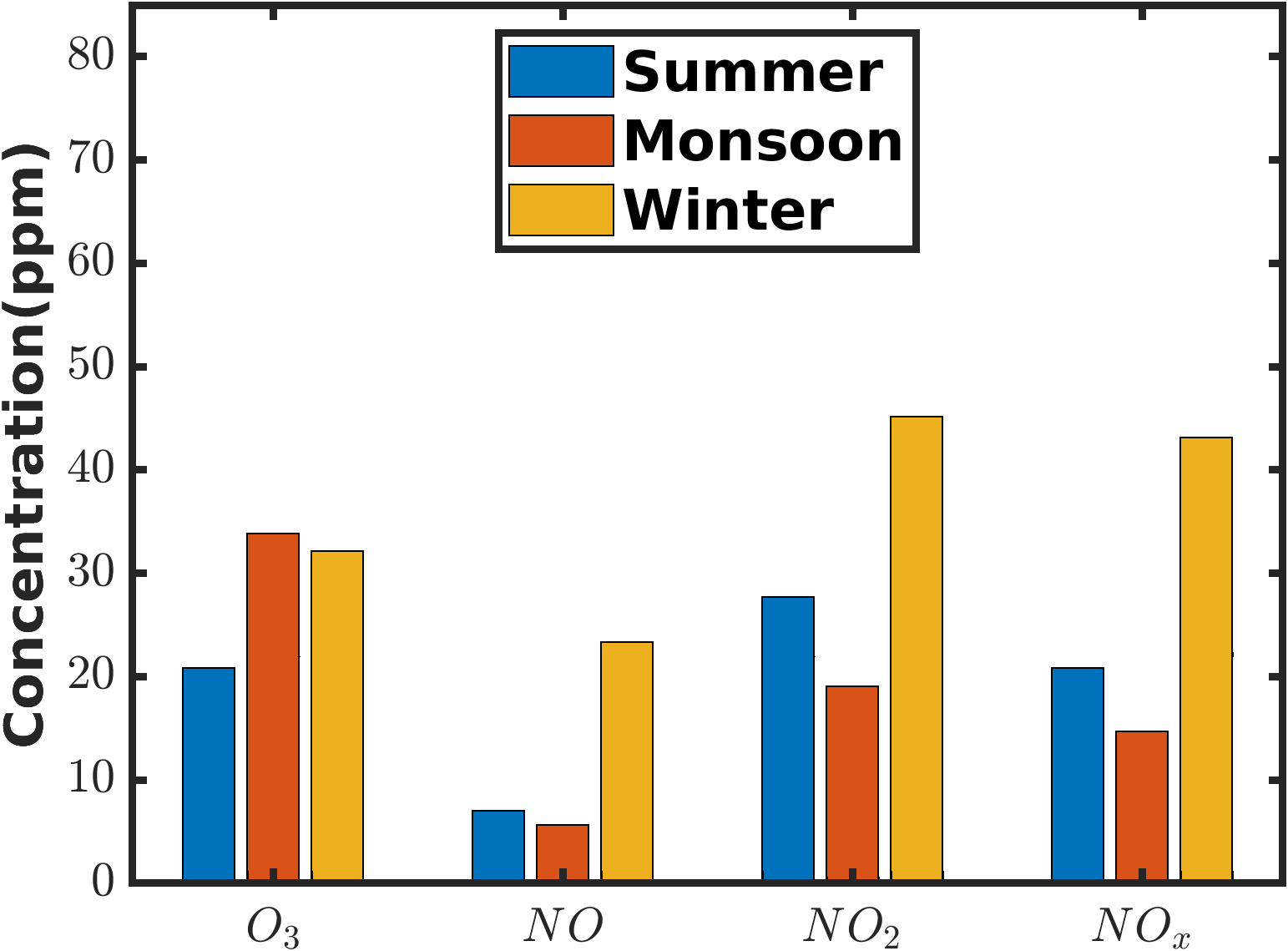}}
  %\hspace*{\fill}
\caption{Concentration of $O_3$, $NO$, $NO_2$, $NO_x$ over a period of $1$ year from August $2020$ to July $2021$ for (a) Commercial and (b) Industrial.}\label{fig2}
\end{figure}
\subsection{Diurnal dependency of $O_3$, $NO$, $NO_2$, and $NO_x$}\label{sec32}
\begin{figure}[H]
\subfloat[]{\label{fig3:a}\includegraphics[width=0.33\textwidth]{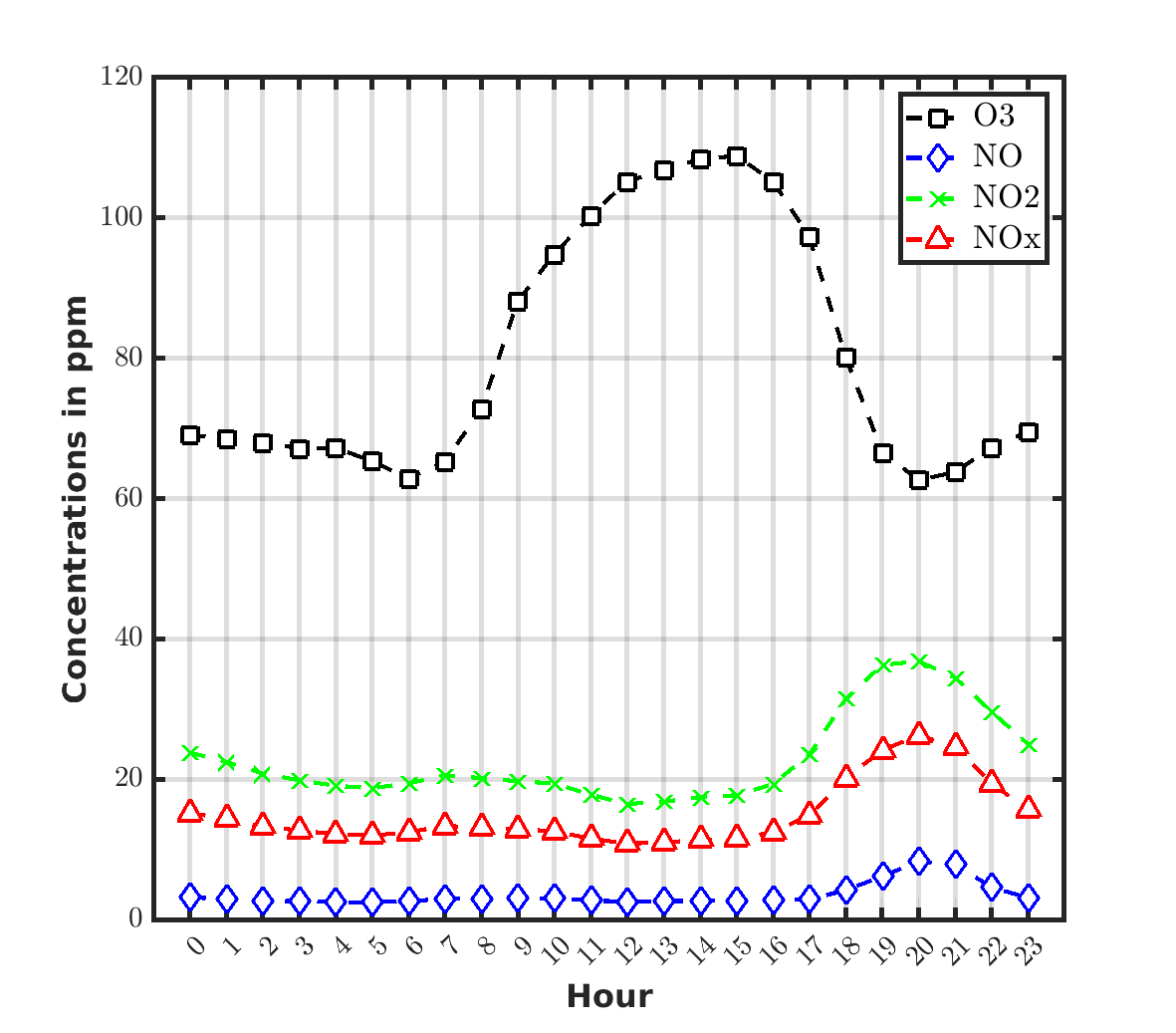}}%
  \subfloat[]{\label{fig3:b}\includegraphics[width=0.33\textwidth]{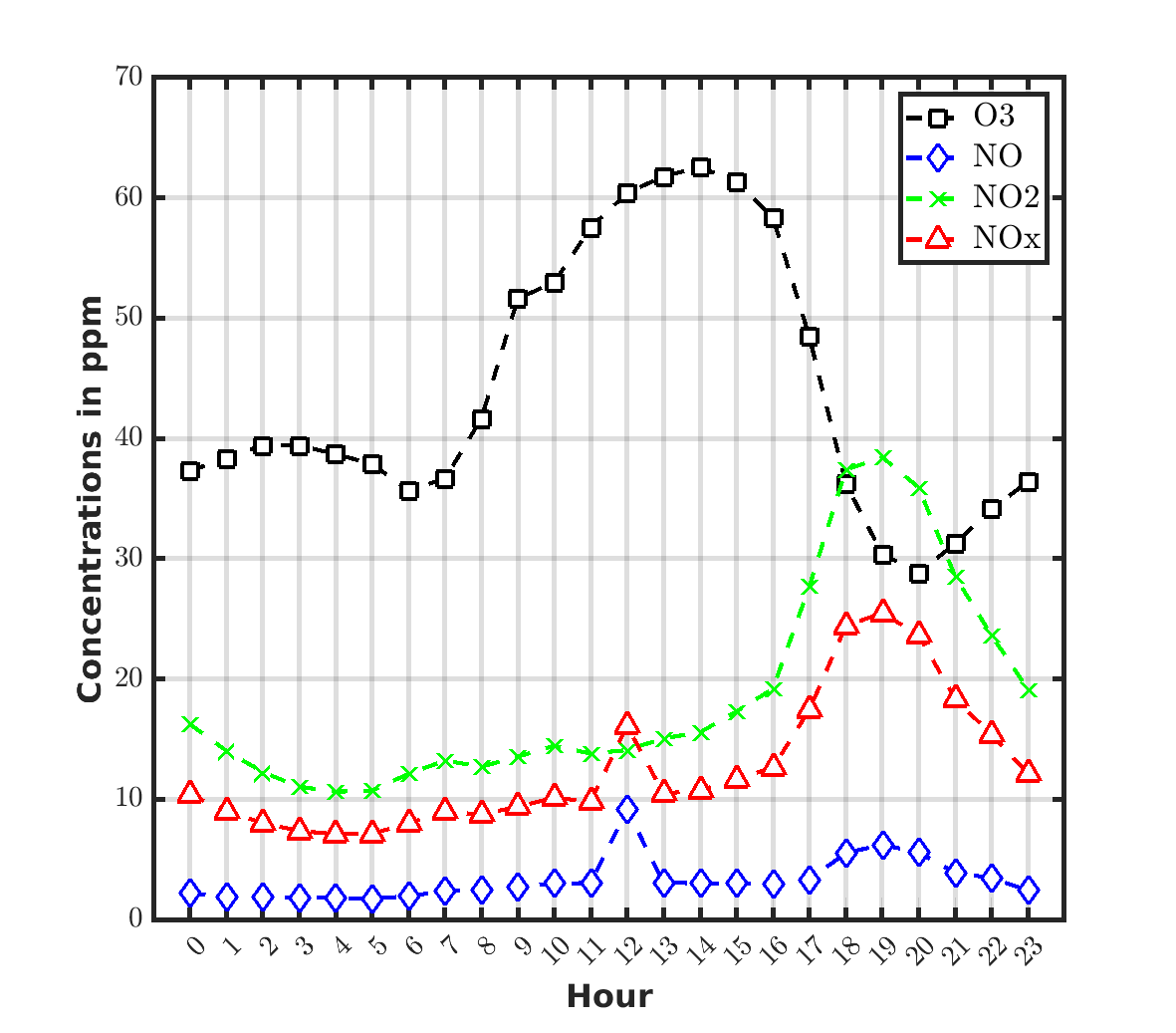}}
  \subfloat[]{\label{fig3:c}\includegraphics[width=0.33\textwidth]{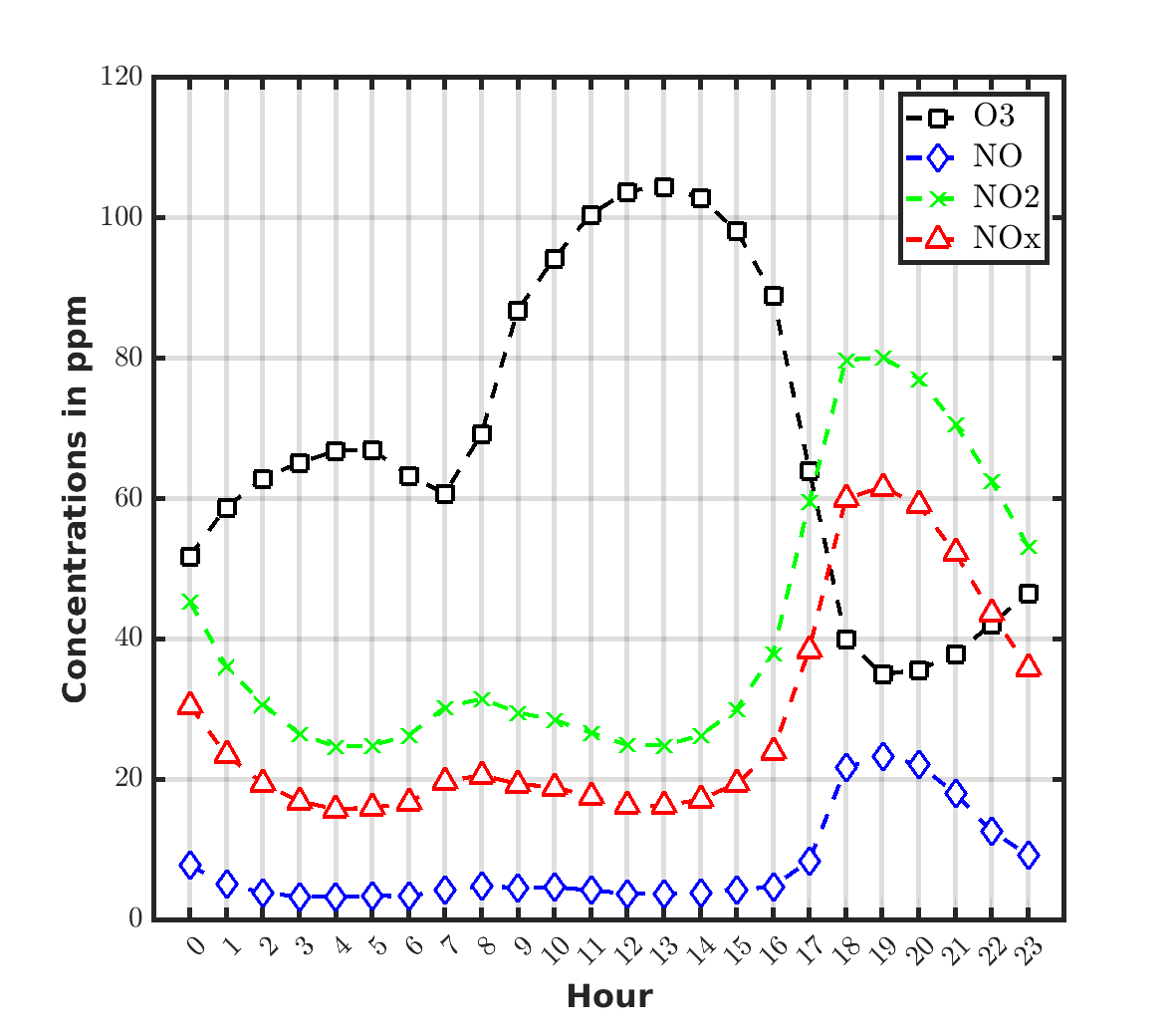}}\\
  \subfloat[]{\label{fig3:d}\includegraphics[width=0.33\textwidth]{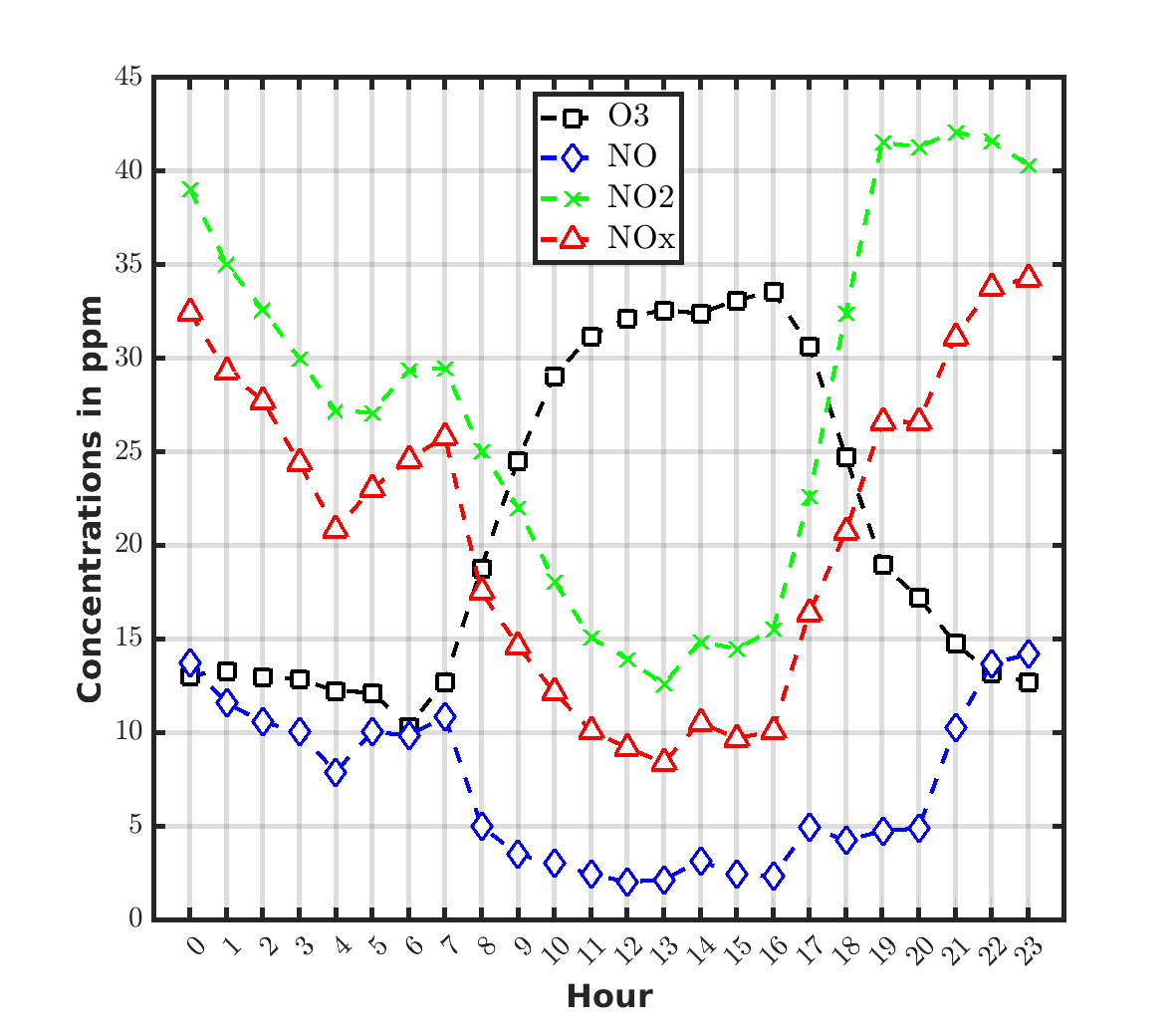}}
  \subfloat[]{\label{fig3:e}\includegraphics[width=0.33\textwidth]{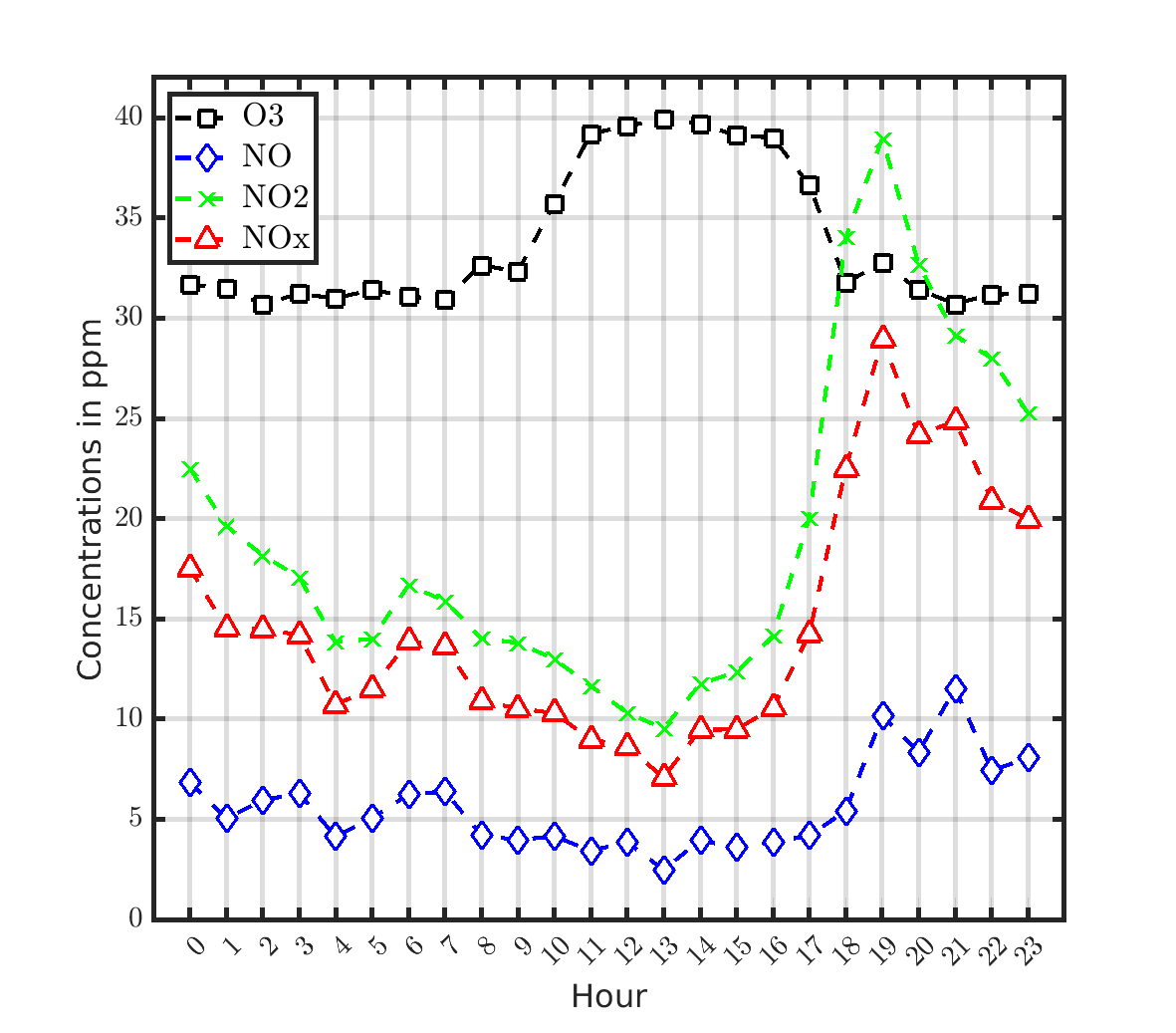}}
    \subfloat[]{\label{fig3:f}\includegraphics[width=0.33\textwidth]{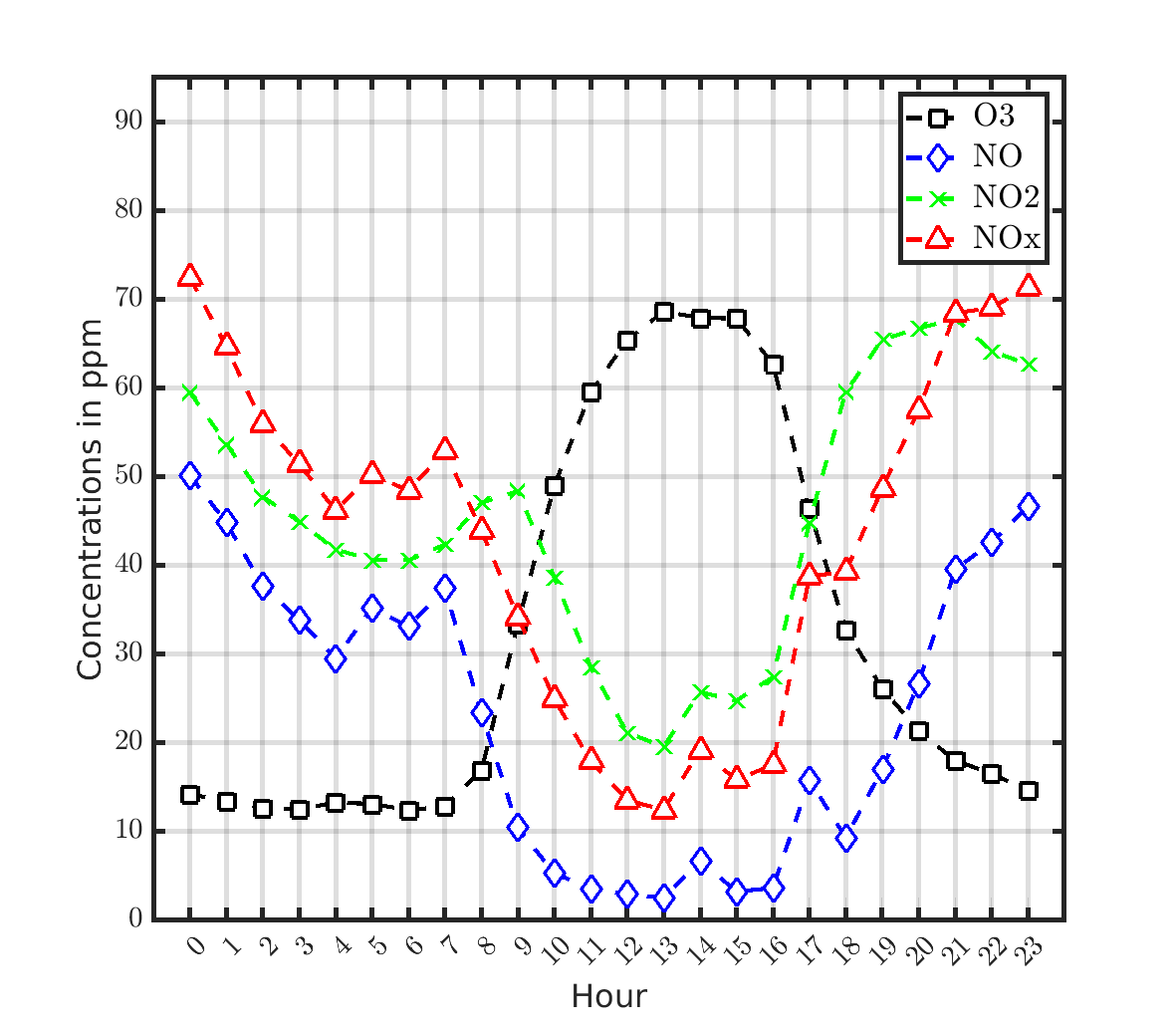}}
  %\hspace*{\fill}
\caption{Hourly concentration of $O_3$, $NO$, $NO_2$, $NO_x$ for Commercial ((a) Summer, (b) Monsoon, (c) Winter) and Industrial ((d) Summer, (e) Monsoon, (f) Winter) areas respectively}\label{fig3}
\end{figure}

Figure~\ref{fig3} shows the averaged hourly variation of $O_3$, $NO$, $NO_2$, $NO_x$ for the entire study period (August $2020$ to July $2021$) for Commercial ((a) Summer, (b) Monsoon, (c) Winter) and Industrial ((d) Summer, (e) Monsoon, (f) Winter) areas, respectively. During all the seasons, the maximum concentration of $NO$ is observed from $18:00$ Hrs to $22:00$ Hrs and  $22:00$ Hrs to $07:00$ Hrs for Commercial and Industrial, respectively, with the probable explanation as increased traffic emissions (Commercial) and reduced night-time boundary layer (Industrial). However, monsoon shows a fluctuation in the trend. $NO$ concentration level decreases after midnight (Commercial) and $07:00$ Hrs (Industrial) in the morning due to its reaction with $O_3$ via its oxidation to $NO_2$ (Reaction-$3$). Within a day, $NO$ goes to its minimum value at around mid-day (for both Industrial and Commercial), which agrees with maximum photolytic phenomena. Trend of $NO_2$ concentration becomes downwards at the early morning time because of photolysis (Reaction-1) which produces $O_3$. After sunset, $NO_2$ concentrations increase and become maximum at around $19:00-20:00$ Hrs (for both industrial and Commercial during all the seasons) in the evening. $O_3$ concentration starts rising after $07:00$ Hrs (Industrial and Commercial) because of enhanced photochemical reactions in the presence of sunlight, and reaches a maximum around $13:00$ Hrs to $15:00$ Hrs (Industrial and Commercial), after which it starts declining again. This behavior also persists throughout the year for different seasons. The opposite nature of $NO_2$ and $O_3$ is because the former leads to the production of $O_3$. The trend, however, is not very clearly visible for Monsoon due to irregularity / high variation in weather.

\subsection{Chemical coupling of $O_3$, $NO$ and $NO_2$}\label{sec33}
\begin{figure}[H]
\subfloat[]{\label{fig3:a}\includegraphics[width=0.45\textwidth]{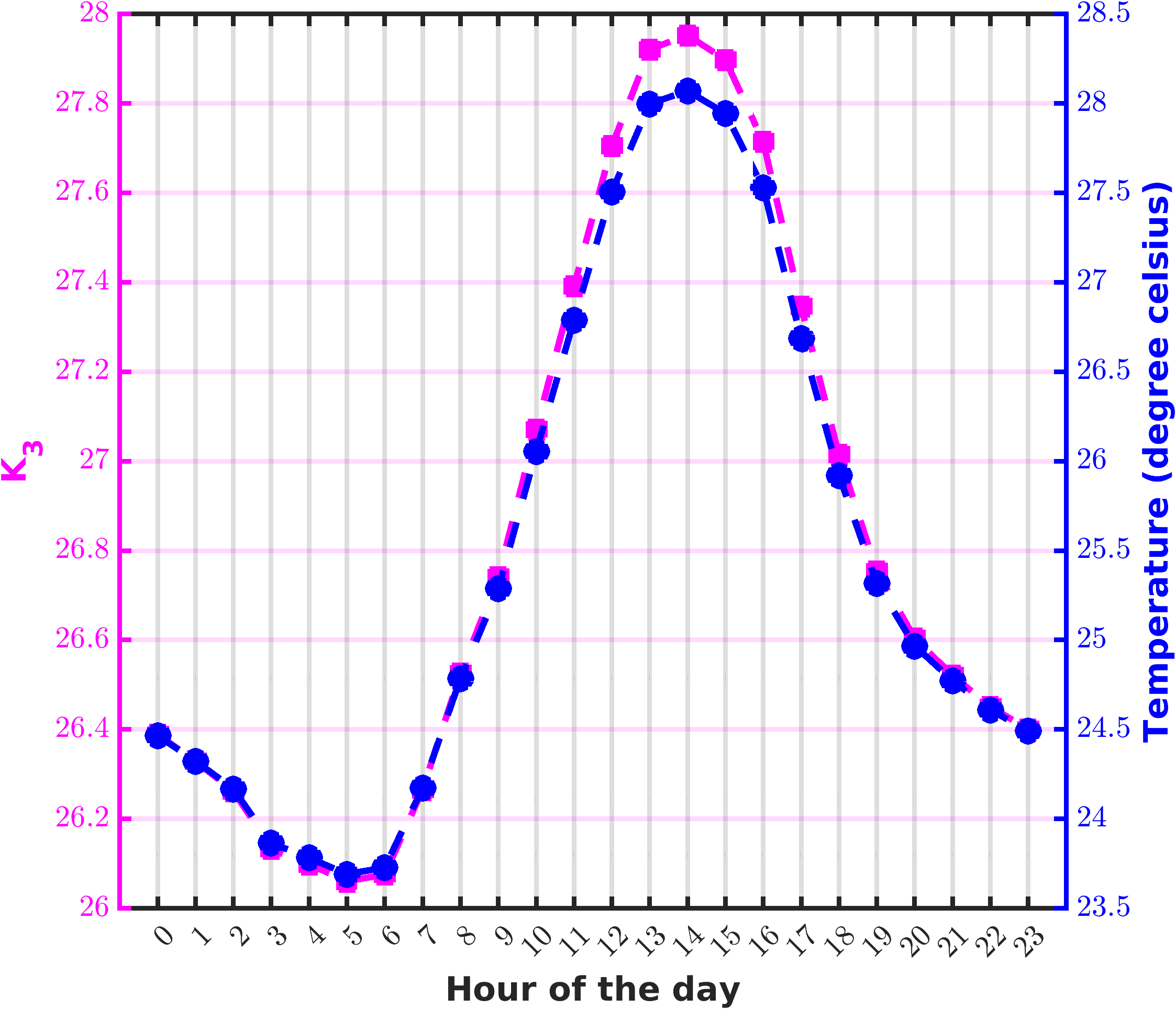}}%
\hspace*{\fill}
\subfloat[]{\label{fig3:b}\includegraphics[width=0.42\textwidth]{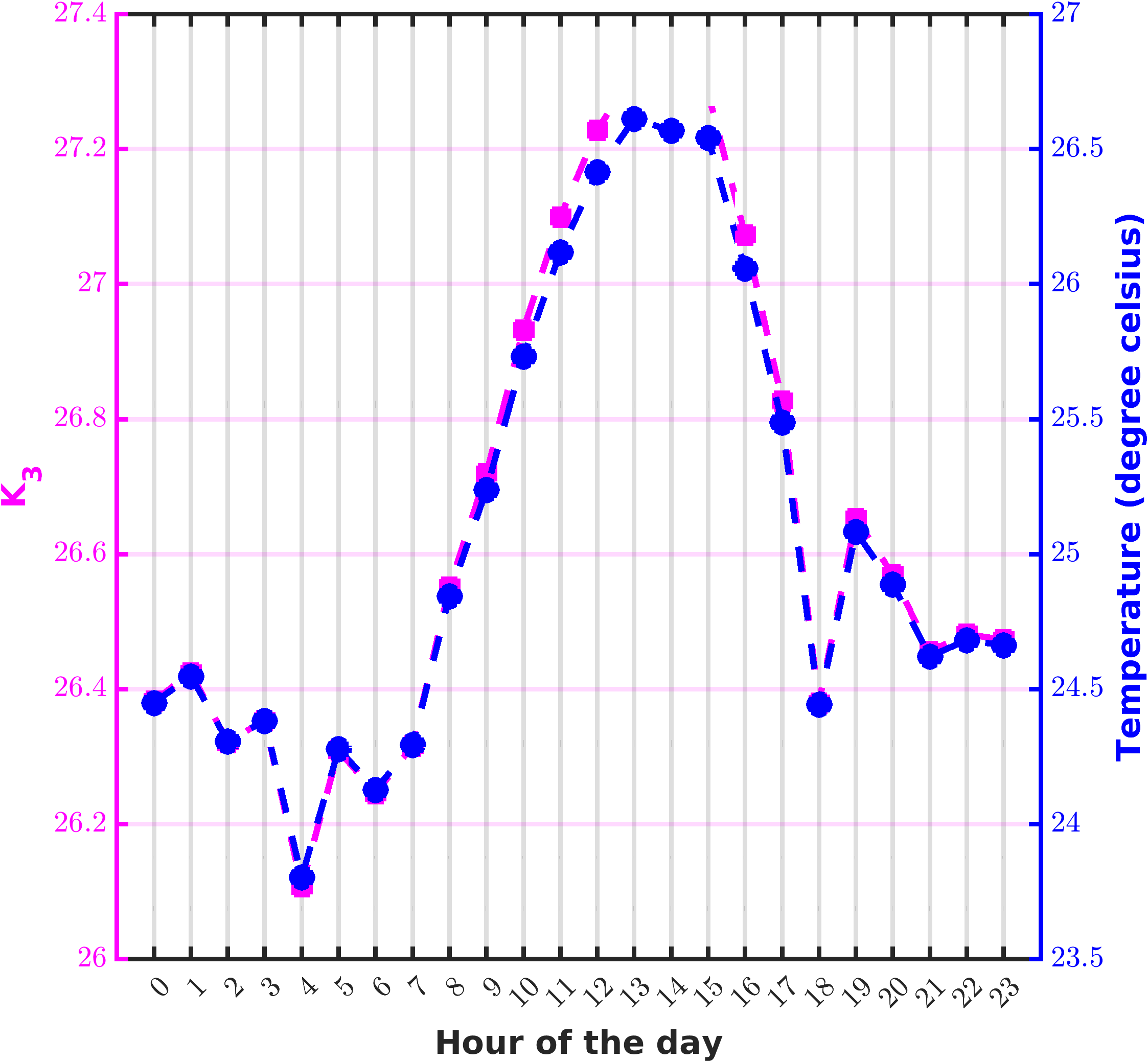}}

\caption{Dependence of $k_3$ on the hour of the day and temperature for (a) Commercial and (b) Industrial area.}\label{fig4}
\end{figure}

Figure~\ref{fig4} shows $k_3$ values against AT. $k_3$ values are calculated using Eq.\ref{eqn10}. $k_3$ increases after $08:00$ Hrs and achieves a peak value of $27-28$ $ppm^{-1} min^{-1}$ at around noon $12:00$ to $15:00$ Hrs, for both the sites (a) Commercial and (b) Industrial, which exactly matches with the occurrence of maximum temperatures in the day. 

\begin{figure}[H]
\subfloat[]{\label{fig5:a}\includegraphics[width=0.33\textwidth]{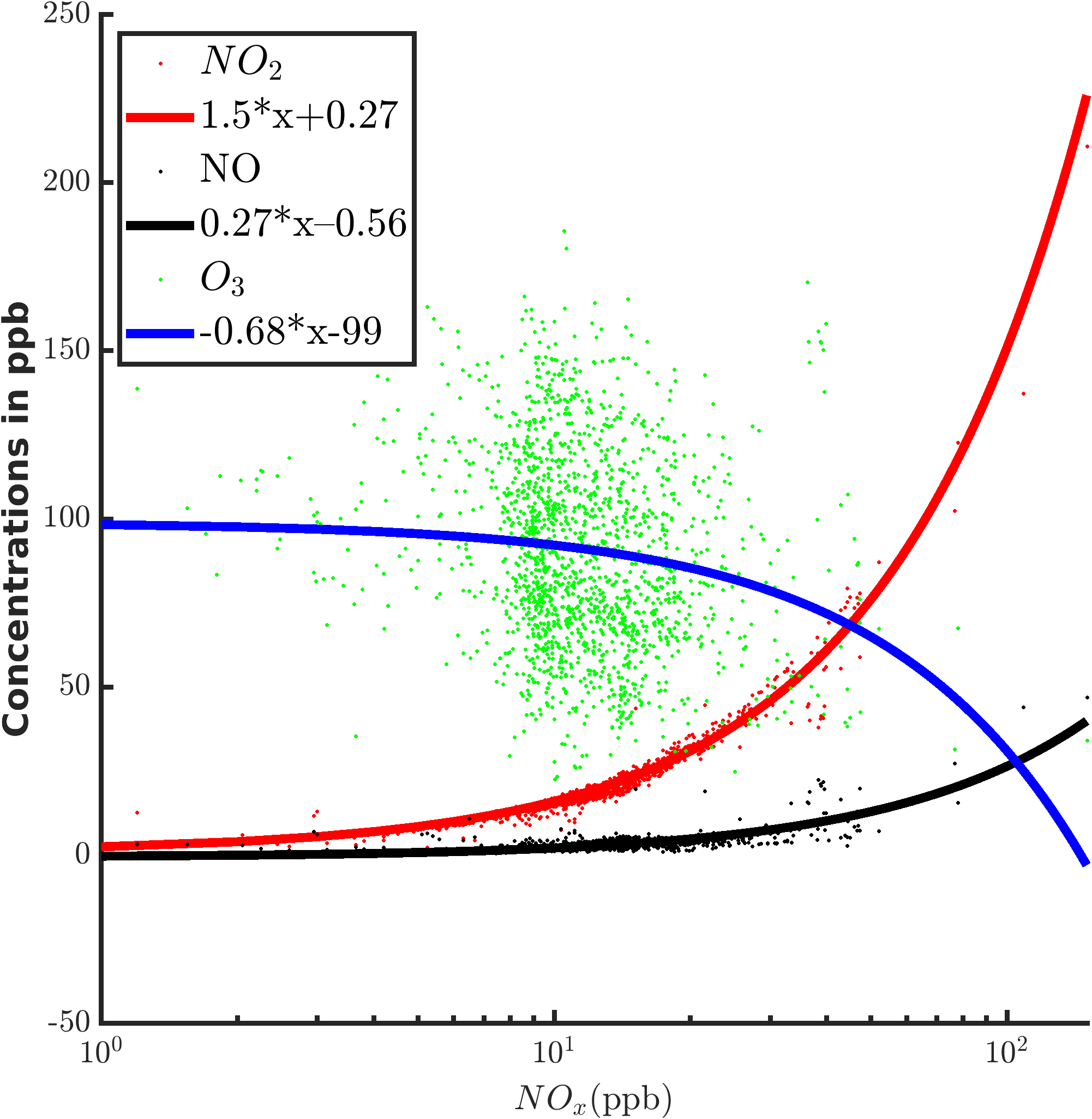}}%
  \subfloat[]{\label{fig5:b}\includegraphics[width=0.33\textwidth]{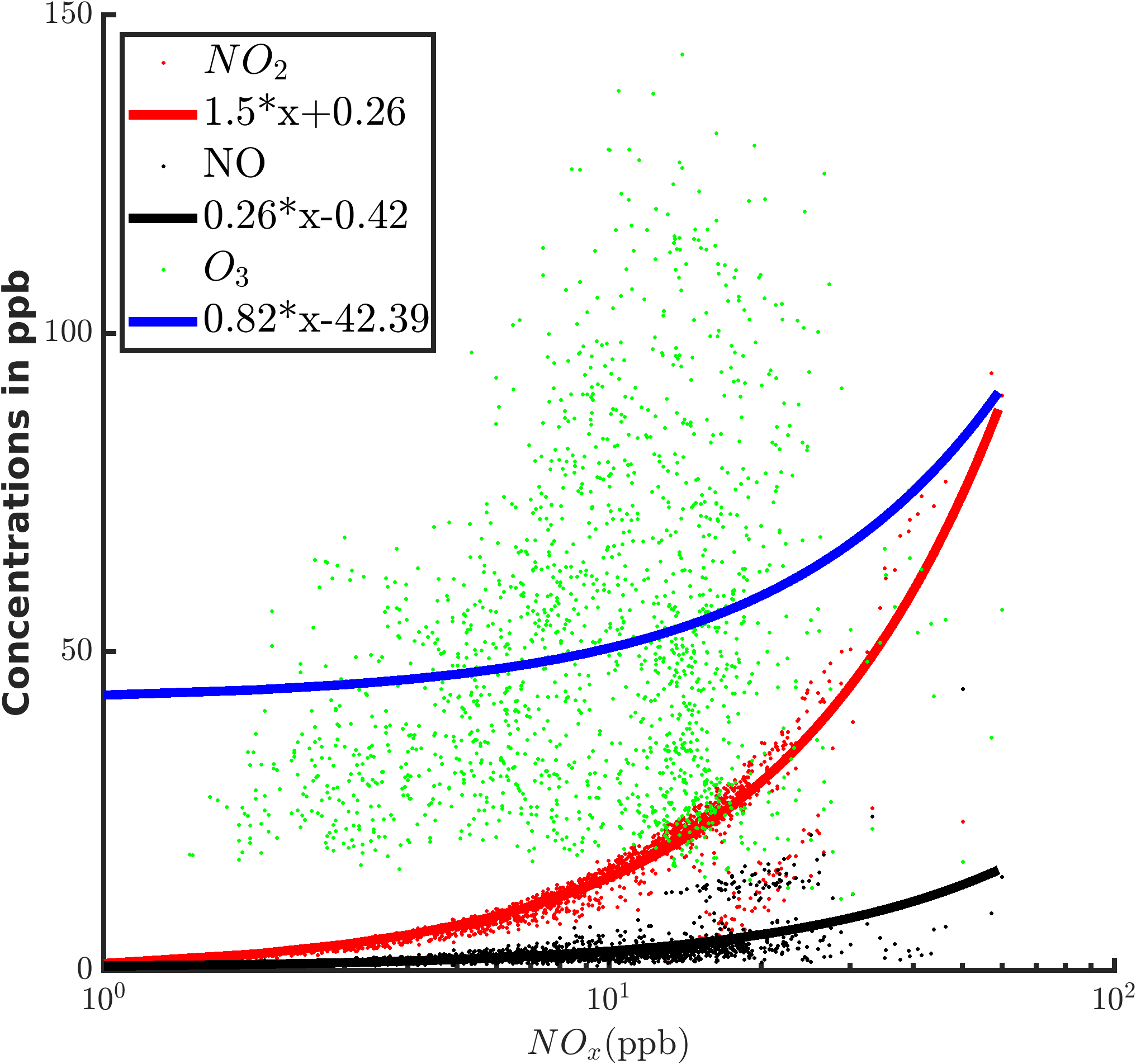}}
  \subfloat[]{\label{fig5:c}\includegraphics[width=0.33\textwidth]{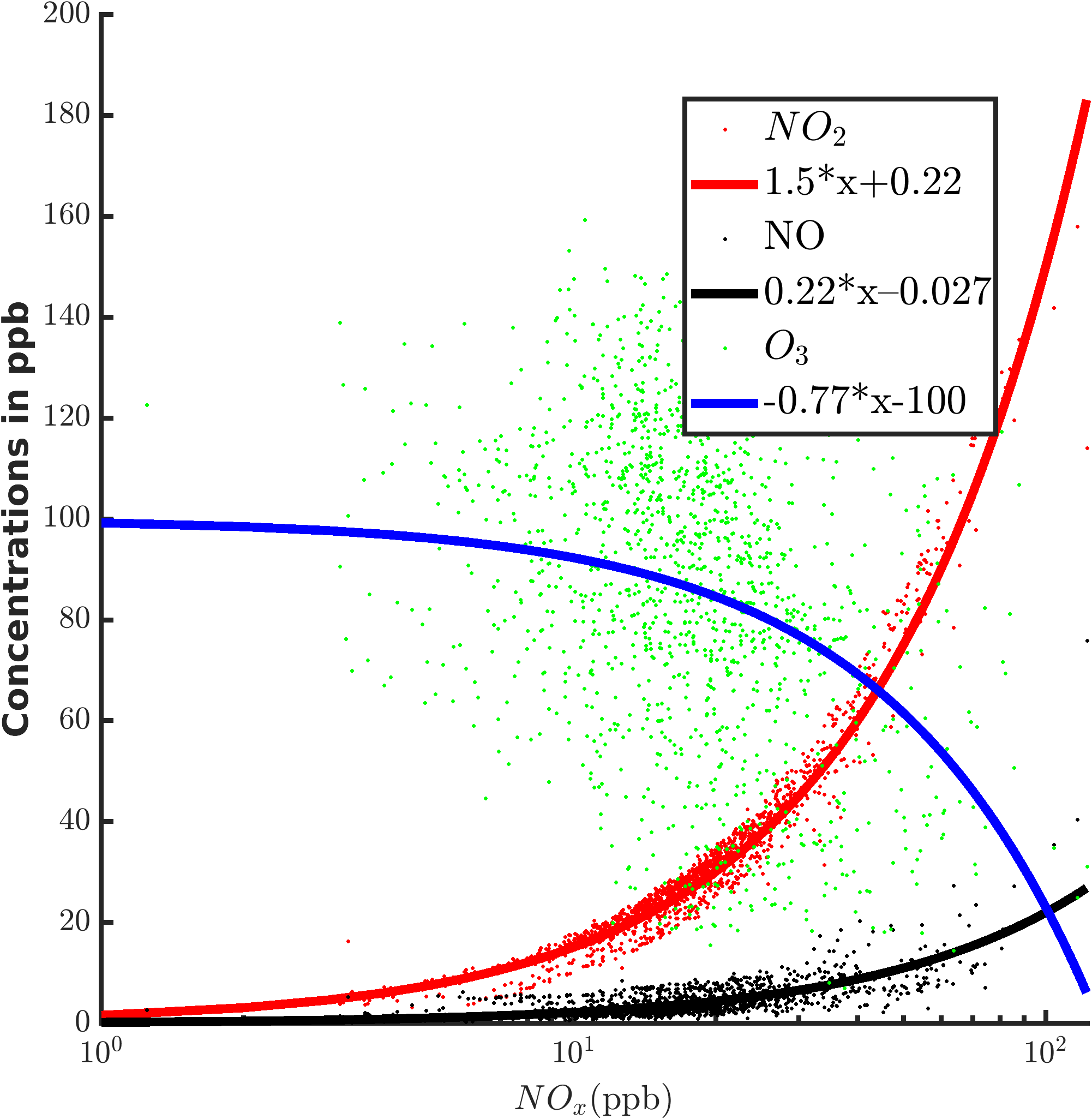}}\\
  \subfloat[]{\label{fig5:d}\includegraphics[width=0.33\textwidth]{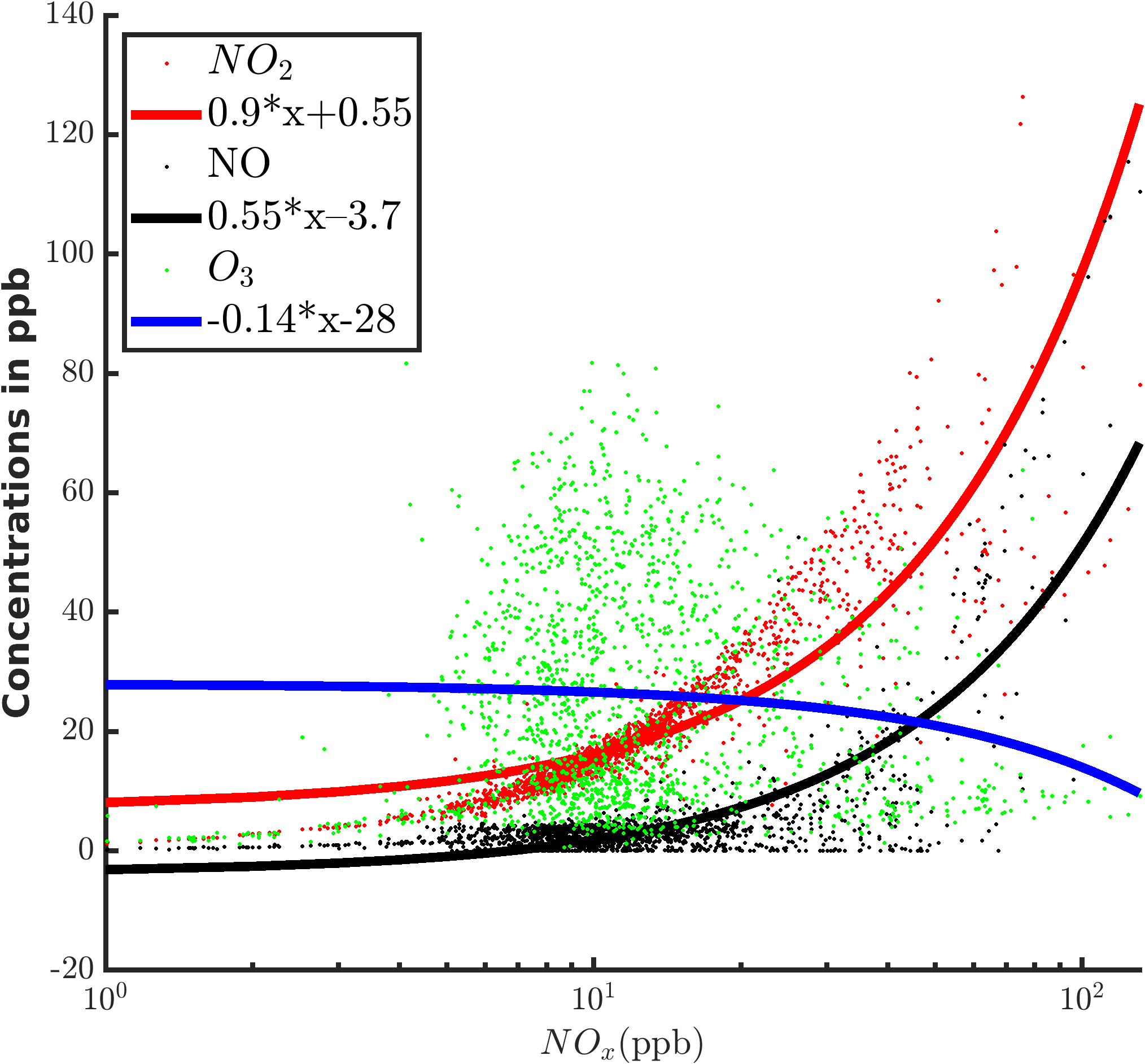}}
  \subfloat[]{\label{fig5:e}\includegraphics[width=0.33\textwidth]{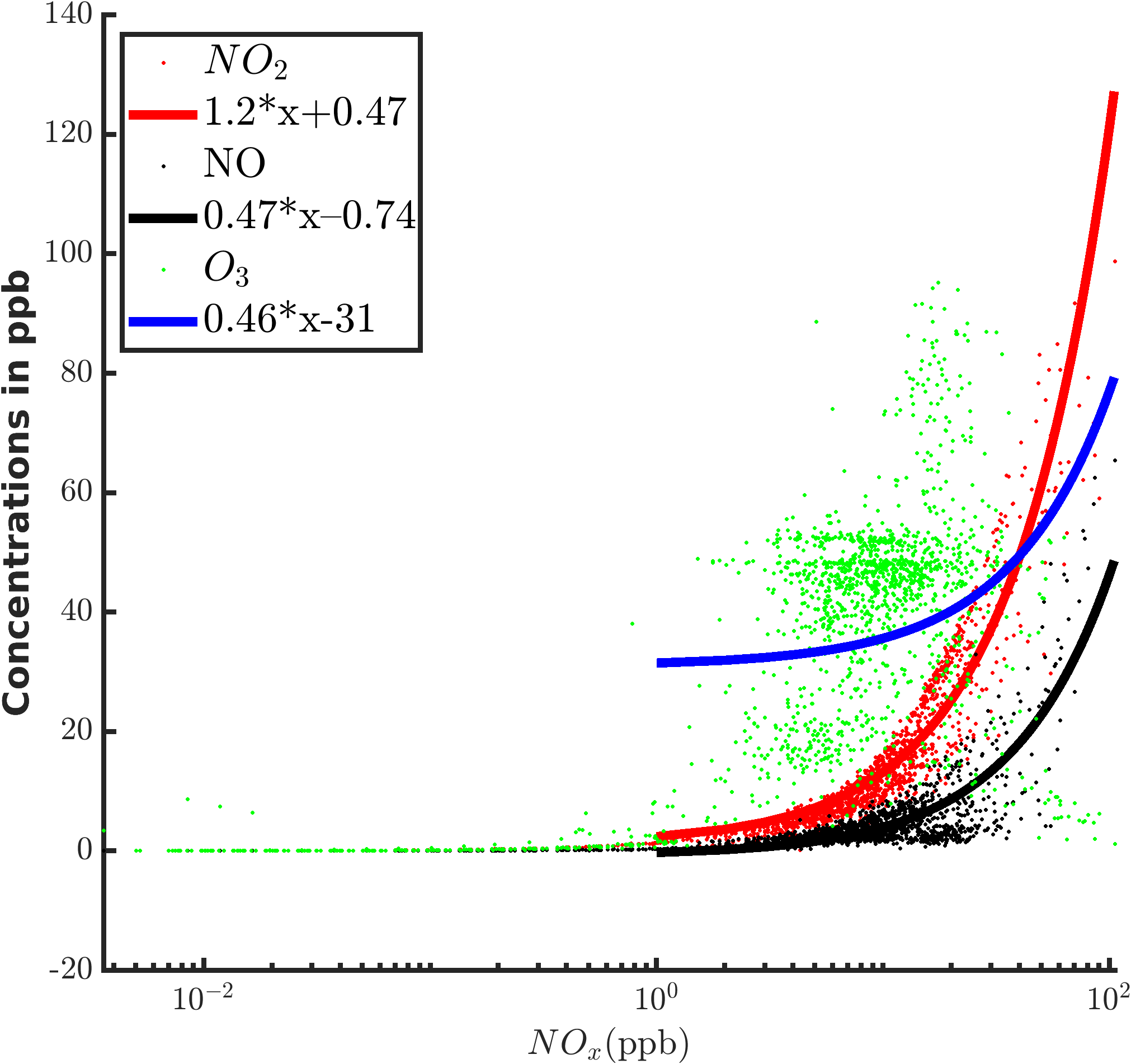}}
    \subfloat[]{\label{fig5:f}\includegraphics[width=0.33\textwidth]{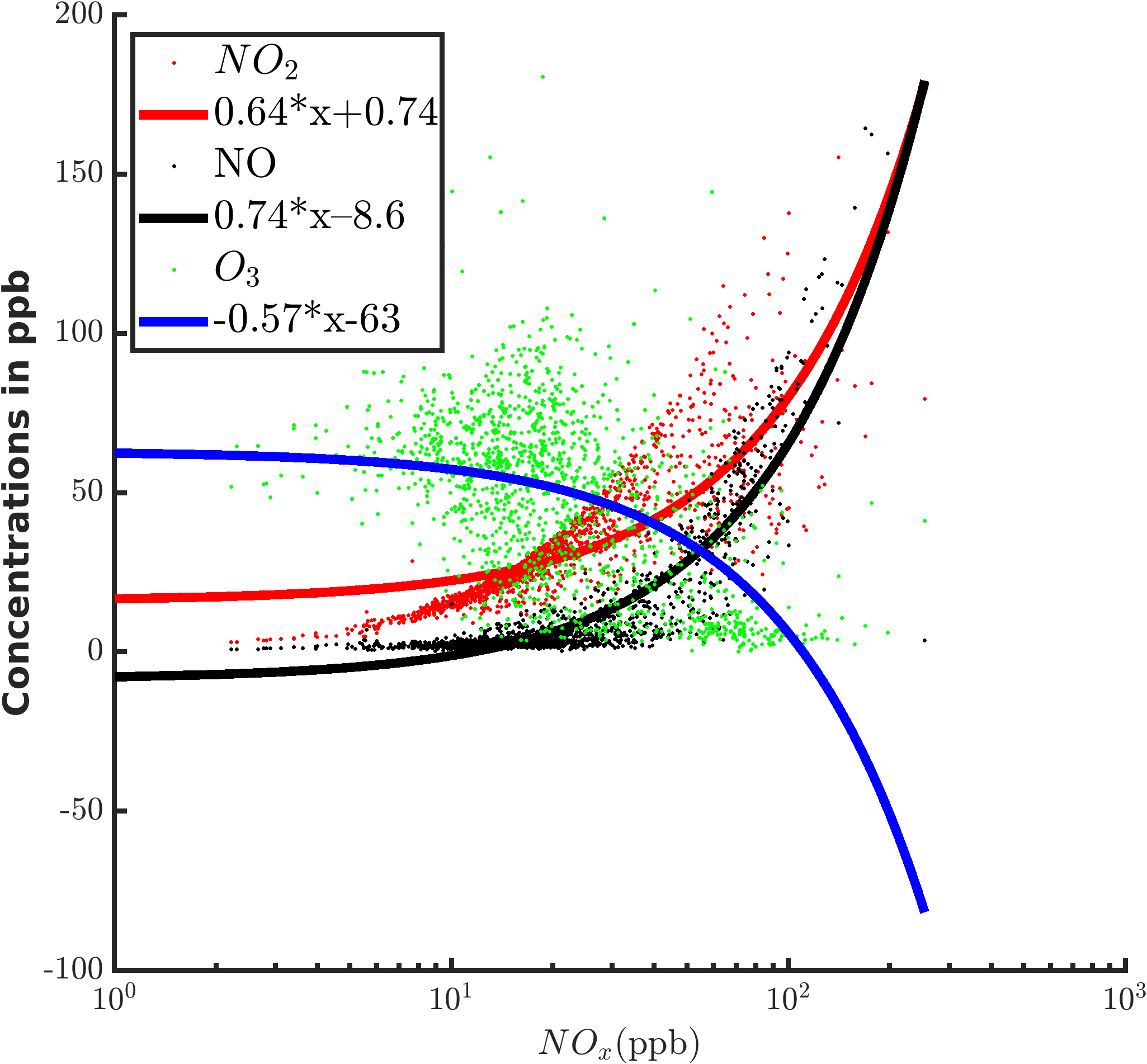}}
  %\hspace*{\fill}
\caption{Chemical coupling of $O_3$, $NO_2$, and $NO$ with $NO_x$ with the daylight concentration for Commercial ((a)Summer, (b)Monsoon, (c)Winter) and Industrial ((d)Summer, (e)Monsoon, (f)Winter) areas respectively}\label{fig5}
\end{figure}

Figure~\ref{fig5} shows the relationship between the daytime concentrations of $O_3$, $NO$, and $NO_2$ versus $NO_x$. The plotted data for $O_3$, $NO$, $NO_2$, and $NO_x$ can be adequately fitted by polynomial curves which proves the presence of a strong chemical coupling between these pollutants, meaning that the concentrations of these pollutants are interdependent and are likely to be affected by chemical reactions occurring within the system. $O_3$ concentration starts to decrease very fast at higher $NO_x$ concentrations; on the other hand, concentrations of $NO$ and $NO_2$ keep increasing at higher $NO_x$ concentration, except for monsoon, which is obvious due to very rapid temperature fluctuation. $O_3$ and $NO$ curves intersect at $102-104$ ppb (Commercial) and $46-55$ ppb (Industrial) $NO_x$. It is observed that $O_3$ concentrations are higher at lower concentrations of $NO_x$ and decrease as $NO_x$ concentrations increase. This is because $O_3$ is formed through a chemical reaction involving UV light, oxygen, and VOCs, which are produced by the burning of fossil fuels and other industrial processes.   As the concentration of $NO_x$ increases, it can interfere with the chemical reactions that produce $O_3$, leading to a decrease in $O_3$ concentrations. At the same time, the increase in $NO_x$ can lead to an increase in the production of $NO$, which can contribute to the formation of $O_3$. $NO_2$ and $O_3$ curves intersect around $44-45$ ppb (Commercial) and $20-39$ ppb (Industrial) of $NO_x$. It clearly shows the chemical coupling between the pollutants. As also explained by the (Reaction-$1$) to (Reaction-$3$), $NO$ and $O_3$ reacts to produce $NO_2$ (Reaction-$3$) which helps in decreasing the concentration of $O_3$ from the environment and makes $NO_2$ dominant at higher concentrations of $NO_x$ \cite{clapp01, mazzeo05, han11, tiwari2015, hassan13}.

\subsection{Separation of local and regional Oxidant contributions}\label{sec34}
\begin{figure}[H]
\subfloat[]{\label{fig3:a}\includegraphics[width=0.45\textwidth]{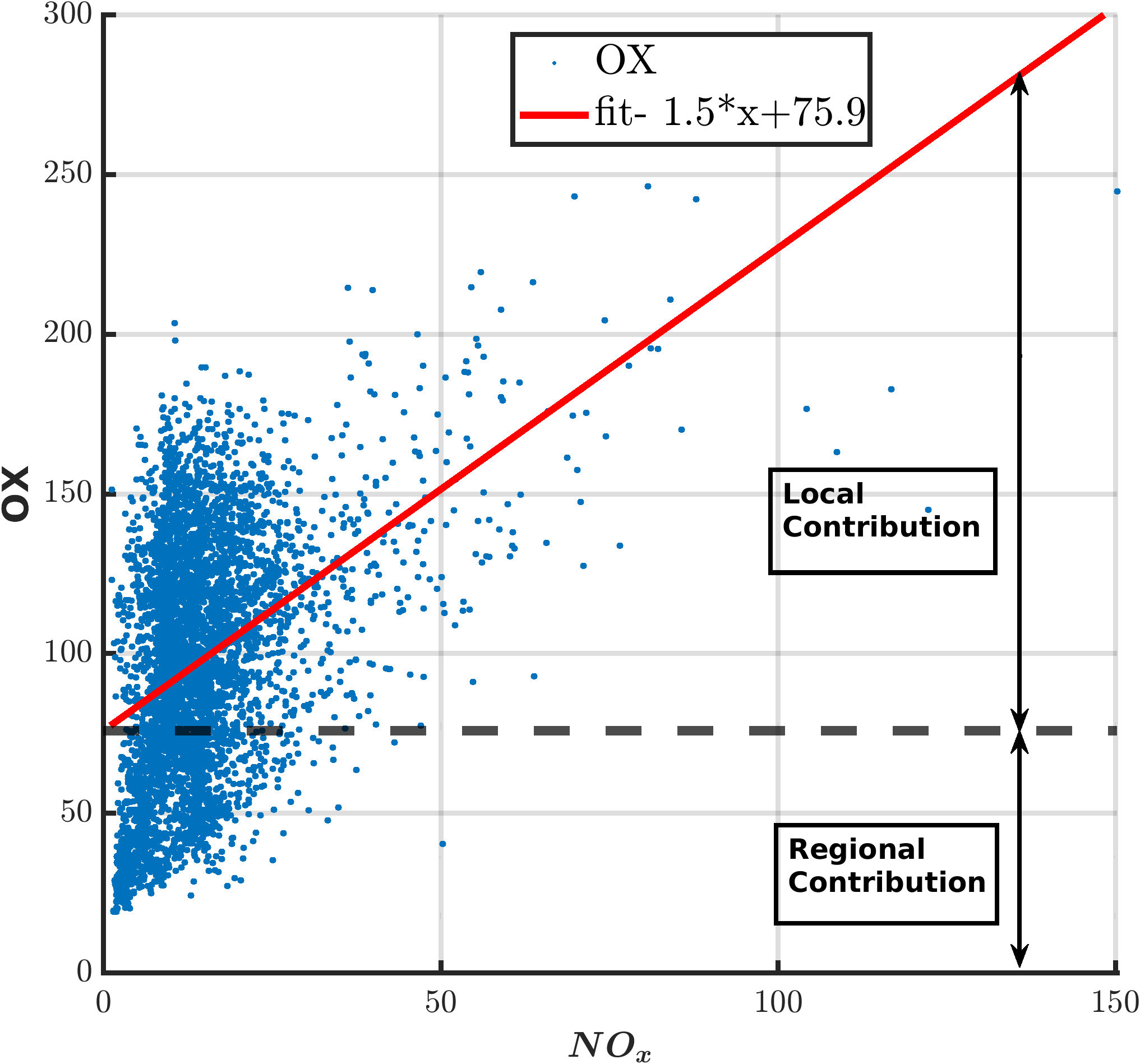}}%
\hspace*{\fill}
\subfloat[]{\label{fig3:b}\includegraphics[width=0.45\textwidth]{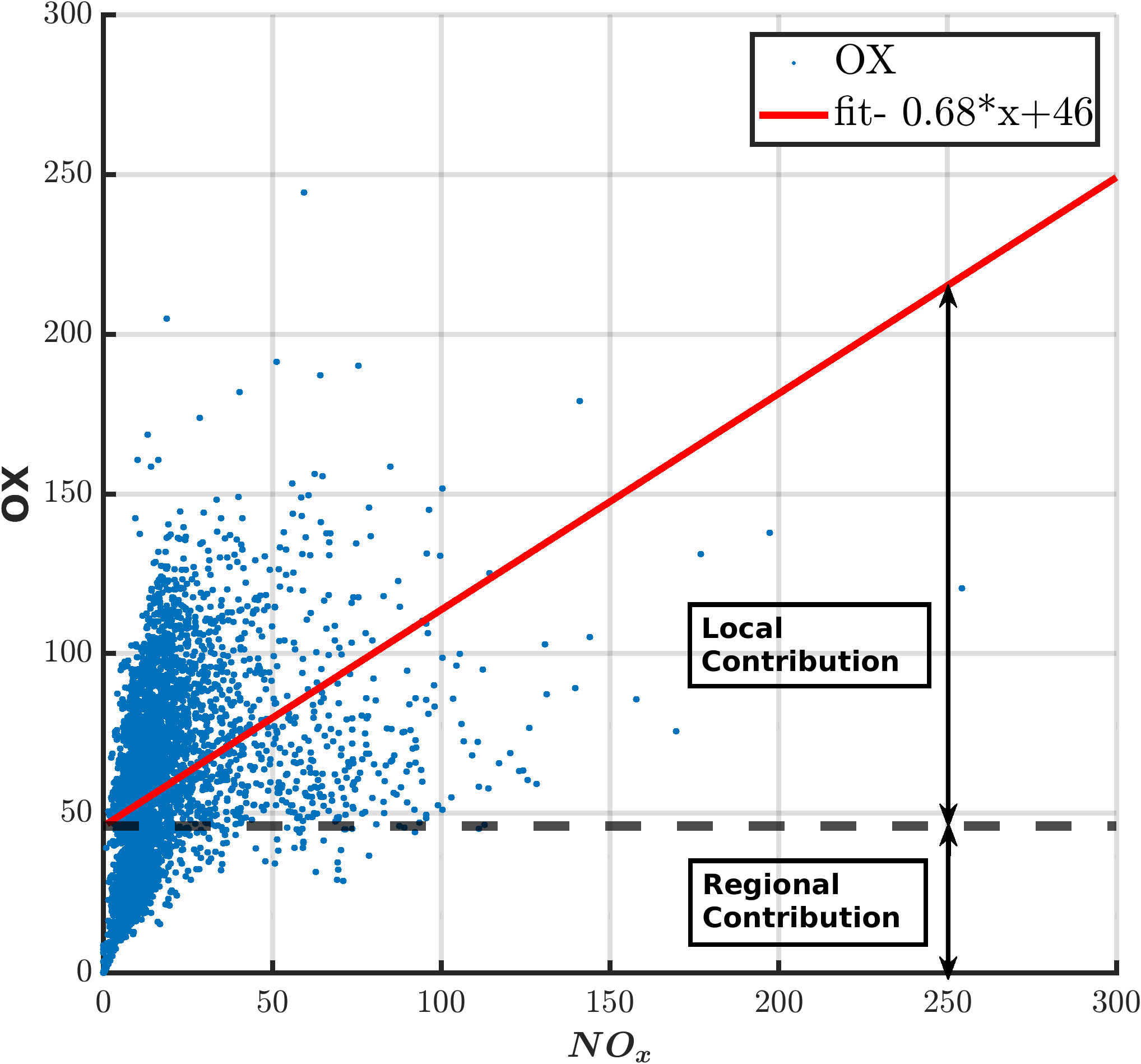}}
  %\hspace*{\fill}
\caption{Variation of daylight values of $[OX]$ with $NO_x$ for (a) Commercial and (b) Industrial.}\label{fig6}
\end{figure}

To analyze the chemical coupling between $O_3$, $NO$, and $NO_2$, we plot the daylight concentrations of the oxidant levels $(OX =O_3+NO_2)$ and $NO_x$ in Figure~\ref{fig6}. The analysis of the chemical coupling between $O_3$, $NO$, and $NO_2$ was done for the winter season because of the presence of atmospheric stability during that time. From the linear regression fit, as mentioned in the inset, we found that for Commercial, the local contribution is more than the regional or background contribution due to increased emissions from mobile sources, even at night \cite{nagpure13}. On the other hand, for Industrial, regional or background contribution is more than the local contribution, which depends on the regional background of $)_3$ concentration level \cite{reddy12}. These results are consistent with the similar findings reported by Badarinath et al. in $2007$ \cite{badarinath07} with the possible explanation of increment of $NO_2$ concentrations due to crop residue burning in the winter season. 

\subsection{Association of $O_3$, $NO$, $NO_2$ and, $NO_x$ with meteorological parameters }\label{sec35}
Table~\ref{tab3} show the relationship between the
daily concentration of ($O_3$, $NO$, $NO_2$, $NO_x$)
and various meteorological parameters for Commercial and Industrial sites using Pearson correlation coefficient. Presence of significant positive correlations $(>0.5)$ between $NO_2$ and $NO$, $NO_x$ and $NO$, and $NO_x$ and $NO_2$ signify the same source for them and affirms the contribution of mobile sources to pollution.
Oxides of Nitrogen (mainly, $NO$, $NO_2$, $NO_x$) shows negative correlation
with $O_3$. It is obvious because $NO_x$ acts as a precursor of $O_3$ \cite{he12} and confirms the `titration effect'(Reaction-$3$). On the other hand, $O_3$ is positively correlated with solar radiation (SR) and ambient temperature (AT), for both sites. This also leads to negatively correlated behavior between $NO_x$ and ambient temperature as $NO_x$ increases the rate of the $O_3$ production (Reaction 1-3) in
the presence of sunlight and high temperatures \cite{{jacob09}}.  These results agree with the investigations reported for different data sets. \cite{teixeira09, pudasainee06, gaur2014, khoder2009, tian2020, nishanth2012, swamy12, kumar2015}.

\begin{table} [H]
    \caption{Pearson correlation coefficients between averaged hourly concentrations of $O_3$, $NO$, $NO_2$ $NO_x$ and meteorological  
parameters (RH-Relative Humidity, WS-Wind speed, SR- Solar Radiation and, AT-ambient temperature) at (a) Commercial area, and (b) Industrial area}
    \label{tab:someLable}
    \includegraphics[width=\textwidth]{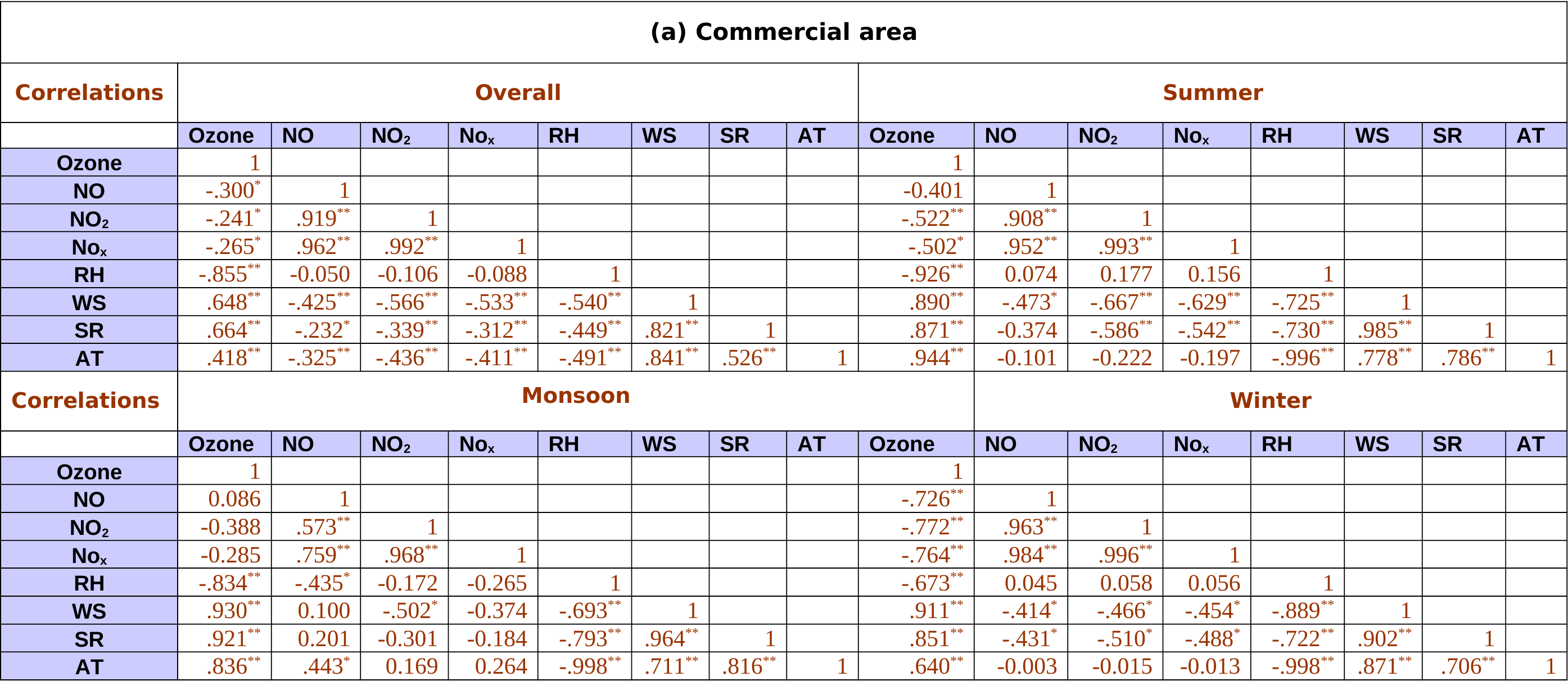}
    \includegraphics[width=\textwidth]{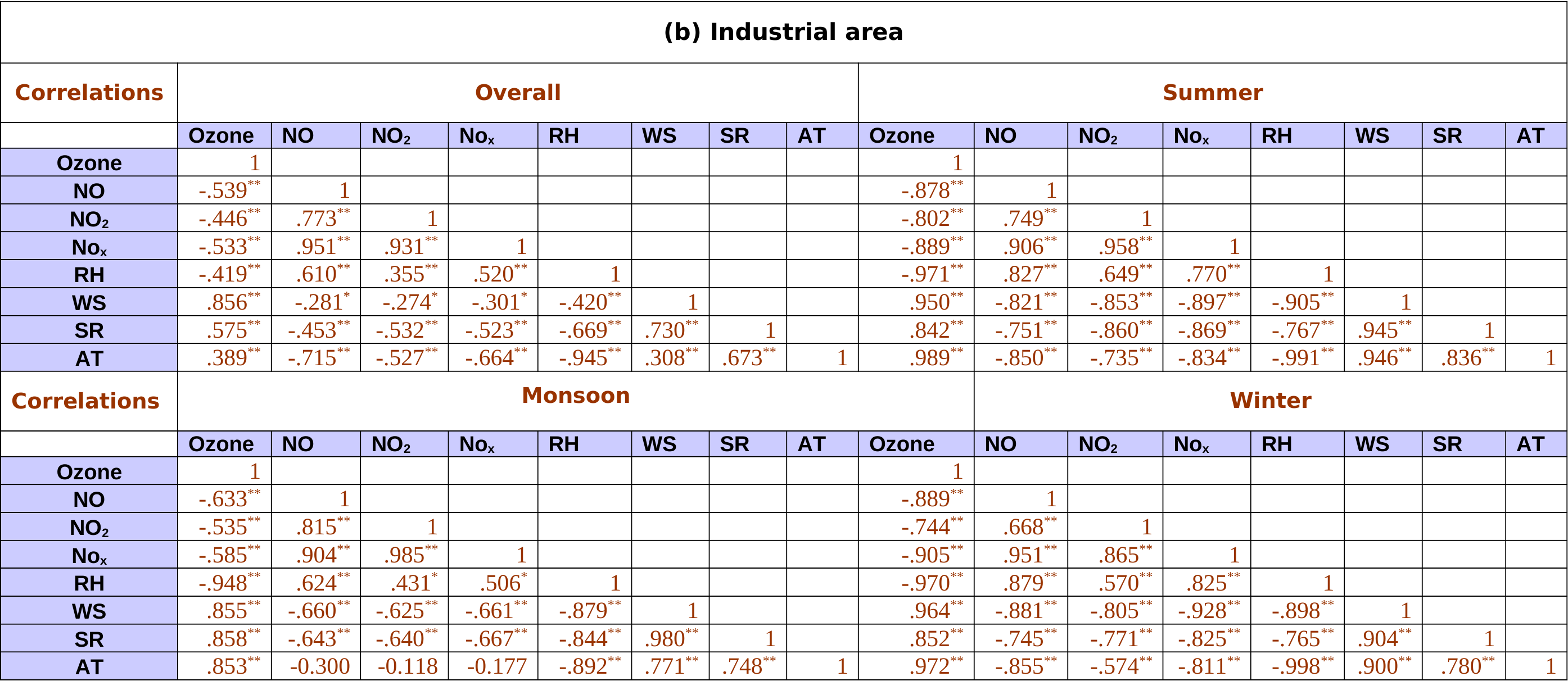}
    \label{tab3}
\end{table}
Wind speed (WS) exhibits a significant correlation with $O_3$, which can be explained as a higher wind speed increases the possibility of dispersion and hence, the mixing of pollutants from nearby sources. Some previous literature also present the similar results \cite{jones10, gasmi2017, korhale2022}.
The correlations of $NO$, $NO_2$, and $NO_x$ with relative humidity (RH) for Commercial and Industrial are inverse in nature. For Commercial, an increase in
relative humidity helps in an updraft of boundary layer air masses, resulting in increased air pollution \cite{elminir05, mavroidis12, ma2000, seinfeld98}. A negative correlation of $O_3$ with relative humidity for both sites is due to the increment of relative humidity, which helps to remove  $O_3$ {reddy12, agudelo13, song11} from ambient air. It should also be noted that very high values of RH, on the other hand, can lead to atmospheric instability and large cloud cover, which declines the rate of photochemical processes and increases the level of $O_3$ by wet deposition \cite{nishanth12}. Our detailed investigation on the association between various pollutants and meteorological parameter thus implies a close relationship between all of them and proves the presence of  multicollinearity.
\section{Conclusion}\label{sec13}
Present investigation considers the continuous measurement of the hourly concentration of surface ozone and its precursors $NO$, $NO_2$, $NO_x$, as well as various meteorological parameters at a commercial site (Bhopal) and an industrial site (Mandideep) at Madhya Pradesh, a central state of India during August $2020$ to July $2021$ to identify trends, patterns, seasonality, spatial variation and their interrelationships. Considering the difference in the profile of the sources at commercial site and at an industrial site, the study anticipated the variance in the magnitude of regional and local contribution of the pollutants. The study revealed that, all the pollutants show very strong seasonality at both the categories of sites viz., commercial and Industrial sites. The diurnal variation of $O_3$ shows an inverse nature to that of $NO$, $NO_2$, $NO_x$ and achieves a peak value when the temperature is maximum and goes to lowest concentration during the night and early morning hours. Chemical coupling among $O_3$, $NO$, and $NO_2$ is also evident and rate coefficient found directly proportional with the ambient temperature. Also, an investigation of variations of daytime concentrations of $NO$, $NO_2$ and $O_3$ with $NO_x$ is performed and the polynomial fitting curve strengthen the concept of chemical coupling between the pollutants. To analyze oxidative dynamics for both the sites, we discuss the $OX$ vs $NO_x$ plot for daytime mean concentration. Slopes and intercepts of the plot determine the local and regional contributions, with higher local contributions for Commercial as compared to Industrial. This can be explained by the difference of the $NO_x$ at both the sites which mainly comes from vehicular exhaust. $O_3$ shows a strong positive correlation with wind speed, temperature and solar radiation except relative humidity. In contrast, $NO_x$ have significant positive correlation with relative humidity and negative correlation with wind speed, temperature and solar radiation.

%\backmatter

\section*{Acknowledgments}
Authors would like to thank Central Pollution Control Board (CPCB), Ministry of Environment, Forests and Climate Change, Government of India for making the data available.  The authors would like to express their gratitude to the the Director, ICMR-NIREH, Bhopal and Vice-Chancellor, VIT Bhopal University for the memorandum of understanding signed between the two institutes, which made this collaboration possible. We also thank Dr. Alf L$\ddot o$effler, Germany and Dr. Thomas H. Seligman, UNAM, Mexico for the useful discussion.

\bibliography{sn-bibliography}% common bib file

%merlin.mbs apsrev4-1.bst 2010-07-25 4.21a (PWD, AO, DPC) hacked
%Control: key (0)
%Control: author (8) initials jnrlst
%Control: editor formatted (1) identically to author
%Control: production of article title (-1) disabled
%Control: page (0) single
%Control: year (1) truncated
%Control: production of eprint (0) enabled
\begin{thebibliography}{62}%
\makeatletter
\providecommand \@ifxundefined [1]{%
 \@ifx{#1\undefined}
}%
\providecommand \@ifnum [1]{%
 \ifnum #1\expandafter \@firstoftwo
 \else \expandafter \@secondoftwo
 \fi
}%
\providecommand \@ifx [1]{%
 \ifx #1\expandafter \@firstoftwo
 \else \expandafter \@secondoftwo
 \fi
}%
\providecommand \natexlab [1]{#1}%
\providecommand \enquote  [1]{``#1''}%
\providecommand \bibnamefont  [1]{#1}%
\providecommand \bibfnamefont [1]{#1}%
\providecommand \citenamefont [1]{#1}%
\providecommand \href@noop [0]{\@secondoftwo}%
\providecommand \href [0]{\begingroup \@sanitize@url \@href}%
\providecommand \@href[1]{\@@startlink{#1}\@@href}%
\providecommand \@@href[1]{\endgroup#1\@@endlink}%
\providecommand \@sanitize@url [0]{\catcode `\\12\catcode `\$12\catcode
  `\&12\catcode `\#12\catcode `\^12\catcode `\_12\catcode `\%12\relax}%
\providecommand \@@startlink[1]{}%
\providecommand \@@endlink[0]{}%
\providecommand \url  [0]{\begingroup\@sanitize@url \@url }%
\providecommand \@url [1]{\endgroup\@href {#1}{\urlprefix }}%
\providecommand \urlprefix  [0]{URL }%
\providecommand \Eprint [0]{\href }%
\providecommand \doibase [0]{http://dx.doi.org/}%
\providecommand \selectlanguage [0]{\@gobble}%
\providecommand \bibinfo  [0]{\@secondoftwo}%
\providecommand \bibfield  [0]{\@secondoftwo}%
\providecommand \translation [1]{[#1]}%
\providecommand \BibitemOpen [0]{}%
\providecommand \bibitemStop [0]{}%
\providecommand \bibitemNoStop [0]{.\EOS\space}%
\providecommand \EOS [0]{\spacefactor3000\relax}%
\providecommand \BibitemShut  [1]{\csname bibitem#1\endcsname}%
\let\auto@bib@innerbib\@empty
%</preamble>
\bibitem [{\citenamefont {Vingarzan}(2004)}]{vin04}%
  \BibitemOpen
  \bibfield  {author} {\bibinfo {author} {\bibfnamefont {R.}~\bibnamefont
  {Vingarzan}},\ }\href {\doibase 10.1016/j.atmosenv.2004.03.030} {\bibfield
  {journal} {\bibinfo  {journal} {ATMOSPHERIC ENVIRONMENT}\ }\textbf {\bibinfo
  {volume} {38}},\ \bibinfo {pages} {3431} (\bibinfo {year}
  {2004})}\BibitemShut {NoStop}%
\bibitem [{\citenamefont {Paoletti}\ \emph {et~al.}(2014)\citenamefont
  {Paoletti}, \citenamefont {De~Marco}, \citenamefont {Beddows}, \citenamefont
  {Harrison},\ and\ \citenamefont {Manning}}]{paoletti14}%
  \BibitemOpen
  \bibfield  {author} {\bibinfo {author} {\bibfnamefont {E.}~\bibnamefont
  {Paoletti}}, \bibinfo {author} {\bibfnamefont {A.}~\bibnamefont {De~Marco}},
  \bibinfo {author} {\bibfnamefont {D.~C.~S.}\ \bibnamefont {Beddows}},
  \bibinfo {author} {\bibfnamefont {R.~M.}\ \bibnamefont {Harrison}}, \ and\
  \bibinfo {author} {\bibfnamefont {W.~J.}\ \bibnamefont {Manning}},\ }\href
  {\doibase 10.1016/j.envpol.2014.04.040} {\bibfield  {journal} {\bibinfo
  {journal} {ENVIRONMENTAL POLLUTION}\ }\textbf {\bibinfo {volume} {192}},\
  \bibinfo {pages} {295} (\bibinfo {year} {2014})}\BibitemShut {NoStop}%
\bibitem [{\citenamefont {Zhang}\ and\ \citenamefont {Oanh}(2002)}]{zhang02}%
  \BibitemOpen
  \bibfield  {author} {\bibinfo {author} {\bibfnamefont {B.}~\bibnamefont
  {Zhang}}\ and\ \bibinfo {author} {\bibfnamefont {N.}~\bibnamefont {Oanh}},\
  }\href {\doibase 10.1016/S1352-2310(02)00348-5} {\bibfield  {journal}
  {\bibinfo  {journal} {ATMOSPHERIC ENVIRONMENT}\ }\textbf {\bibinfo {volume}
  {36}},\ \bibinfo {pages} {4211} (\bibinfo {year} {2002})}\BibitemShut
  {NoStop}%
\bibitem [{\citenamefont {Geddes}\ \emph {et~al.}(2009)\citenamefont {Geddes},
  \citenamefont {Murphy},\ and\ \citenamefont {Wang}}]{geddes09}%
  \BibitemOpen
  \bibfield  {author} {\bibinfo {author} {\bibfnamefont {J.~A.}\ \bibnamefont
  {Geddes}}, \bibinfo {author} {\bibfnamefont {J.~G.}\ \bibnamefont {Murphy}},
  \ and\ \bibinfo {author} {\bibfnamefont {D.~K.}\ \bibnamefont {Wang}},\
  }\href {\doibase 10.1016/j.atmosenv.2009.03.053} {\bibfield  {journal}
  {\bibinfo  {journal} {ATMOSPHERIC ENVIRONMENT}\ }\textbf {\bibinfo {volume}
  {43}},\ \bibinfo {pages} {3407} (\bibinfo {year} {2009})}\BibitemShut
  {NoStop}%
\bibitem [{\citenamefont {Dhanya}\ \emph {et~al.}(2021)\citenamefont {Dhanya},
  \citenamefont {Pranesha}, \citenamefont {Nagaraja}, \citenamefont {Chate},\
  and\ \citenamefont {Beig}}]{dhanya2021}%
  \BibitemOpen
  \bibfield  {author} {\bibinfo {author} {\bibfnamefont {G.}~\bibnamefont
  {Dhanya}}, \bibinfo {author} {\bibfnamefont {T.}~\bibnamefont {Pranesha}},
  \bibinfo {author} {\bibfnamefont {K.}~\bibnamefont {Nagaraja}}, \bibinfo
  {author} {\bibfnamefont {D.}~\bibnamefont {Chate}}, \ and\ \bibinfo {author}
  {\bibfnamefont {G.}~\bibnamefont {Beig}},\ }\href@noop {} {\bibfield
  {journal} {\bibinfo  {journal} {Environmental Monitoring and Assessment}\
  }\textbf {\bibinfo {volume} {193}},\ \bibinfo {pages} {1} (\bibinfo {year}
  {2021})}\BibitemShut {NoStop}%
\bibitem [{\citenamefont {Shukla}\ \emph {et~al.}(2021)\citenamefont {Shukla},
  \citenamefont {Dadheech}, \citenamefont {Kumar},\ and\ \citenamefont
  {Khare}}]{shukla2021}%
  \BibitemOpen
  \bibfield  {author} {\bibinfo {author} {\bibfnamefont {K.}~\bibnamefont
  {Shukla}}, \bibinfo {author} {\bibfnamefont {N.}~\bibnamefont {Dadheech}},
  \bibinfo {author} {\bibfnamefont {P.}~\bibnamefont {Kumar}}, \ and\ \bibinfo
  {author} {\bibfnamefont {M.}~\bibnamefont {Khare}},\ }\href@noop {}
  {\bibfield  {journal} {\bibinfo  {journal} {Chemosphere}\ }\textbf {\bibinfo
  {volume} {272}},\ \bibinfo {pages} {129611} (\bibinfo {year}
  {2021})}\BibitemShut {NoStop}%
\bibitem [{\citenamefont {Notario}\ \emph {et~al.}(2012)\citenamefont
  {Notario}, \citenamefont {Bravo}, \citenamefont {Antonio~Adame},
  \citenamefont {Diaz-de Mera}, \citenamefont {Aranda}, \citenamefont
  {Rodriguez},\ and\ \citenamefont {Rodriguez}}]{notario12}%
  \BibitemOpen
  \bibfield  {author} {\bibinfo {author} {\bibfnamefont {A.}~\bibnamefont
  {Notario}}, \bibinfo {author} {\bibfnamefont {I.}~\bibnamefont {Bravo}},
  \bibinfo {author} {\bibfnamefont {J.}~\bibnamefont {Antonio~Adame}}, \bibinfo
  {author} {\bibfnamefont {Y.}~\bibnamefont {Diaz-de Mera}}, \bibinfo {author}
  {\bibfnamefont {A.}~\bibnamefont {Aranda}}, \bibinfo {author} {\bibfnamefont
  {A.}~\bibnamefont {Rodriguez}}, \ and\ \bibinfo {author} {\bibfnamefont
  {D.}~\bibnamefont {Rodriguez}},\ }\href {\doibase
  10.1016/j.atmosres.2011.10.008} {\bibfield  {journal} {\bibinfo  {journal}
  {ATMOSPHERIC RESEARCH}\ }\textbf {\bibinfo {volume} {104}},\ \bibinfo {pages}
  {217} (\bibinfo {year} {2012})}\BibitemShut {NoStop}%
\bibitem [{\citenamefont {Mahapatra}\ \emph {et~al.}(2014)\citenamefont
  {Mahapatra}, \citenamefont {Panda}, \citenamefont {Walvekar}, \citenamefont
  {Kumar}, \citenamefont {Das},\ and\ \citenamefont {Gurjar}}]{mahapatra2014}%
  \BibitemOpen
  \bibfield  {author} {\bibinfo {author} {\bibfnamefont {P.~S.}\ \bibnamefont
  {Mahapatra}}, \bibinfo {author} {\bibfnamefont {S.}~\bibnamefont {Panda}},
  \bibinfo {author} {\bibfnamefont {P.}~\bibnamefont {Walvekar}}, \bibinfo
  {author} {\bibfnamefont {R.}~\bibnamefont {Kumar}}, \bibinfo {author}
  {\bibfnamefont {T.}~\bibnamefont {Das}}, \ and\ \bibinfo {author}
  {\bibfnamefont {B.~R.}\ \bibnamefont {Gurjar}},\ }\href@noop {} {\bibfield
  {journal} {\bibinfo  {journal} {Environmental Science and Pollution
  Research}\ }\textbf {\bibinfo {volume} {21}},\ \bibinfo {pages} {11418}
  (\bibinfo {year} {2014})}\BibitemShut {NoStop}%
\bibitem [{Note1()}]{Note1}%
  \BibitemOpen
  \bibinfo {note} {Gandhiok, Jasjeev, ``Average Madhya Pradesh resident losing
  $2.9$ years of life due}\BibitemShut {NoStop}%
\bibitem [{Note2()}]{Note2}%
  \BibitemOpen
  \bibinfo {note} {\protect \url {https://aqli.epic.uchicago.edu/}}\BibitemShut
  {NoStop}%
\bibitem [{\citenamefont {Jerrett}\ \emph {et~al.}(2009)\citenamefont
  {Jerrett}, \citenamefont {Burnett}, \citenamefont {Pope}, \citenamefont
  {Ito}, \citenamefont {Thurston}, \citenamefont {Krewski}, \citenamefont
  {Shi}, \citenamefont {Calle},\ and\ \citenamefont {Thun}}]{jerrett09}%
  \BibitemOpen
  \bibfield  {author} {\bibinfo {author} {\bibfnamefont {M.}~\bibnamefont
  {Jerrett}}, \bibinfo {author} {\bibfnamefont {R.~T.}\ \bibnamefont
  {Burnett}}, \bibinfo {author} {\bibfnamefont {C.~A.}\ \bibnamefont {Pope},
  \bibfnamefont {II}}, \bibinfo {author} {\bibfnamefont {K.}~\bibnamefont
  {Ito}}, \bibinfo {author} {\bibfnamefont {G.}~\bibnamefont {Thurston}},
  \bibinfo {author} {\bibfnamefont {D.}~\bibnamefont {Krewski}}, \bibinfo
  {author} {\bibfnamefont {Y.}~\bibnamefont {Shi}}, \bibinfo {author}
  {\bibfnamefont {E.}~\bibnamefont {Calle}}, \ and\ \bibinfo {author}
  {\bibfnamefont {M.}~\bibnamefont {Thun}},\ }\href {\doibase
  10.1056/NEJMoa0803894} {\bibfield  {journal} {\bibinfo  {journal} {NEW
  ENGLAND JOURNAL OF MEDICINE}\ }\textbf {\bibinfo {volume} {360}},\ \bibinfo
  {pages} {1085} (\bibinfo {year} {2009})}\BibitemShut {NoStop}%
\bibitem [{\citenamefont {Kim}\ \emph {et~al.}(2020)\citenamefont {Kim},
  \citenamefont {Kim},\ and\ \citenamefont {Kim}}]{kim2020}%
  \BibitemOpen
  \bibfield  {author} {\bibinfo {author} {\bibfnamefont {S.-Y.}\ \bibnamefont
  {Kim}}, \bibinfo {author} {\bibfnamefont {E.}~\bibnamefont {Kim}}, \ and\
  \bibinfo {author} {\bibfnamefont {W.~J.}\ \bibnamefont {Kim}},\ }\href@noop
  {} {\bibfield  {journal} {\bibinfo  {journal} {Tuberculosis and Respiratory
  Diseases}\ }\textbf {\bibinfo {volume} {83}},\ \bibinfo {pages} {S6}
  (\bibinfo {year} {2020})}\BibitemShut {NoStop}%
\bibitem [{\citenamefont {Rajak}\ and\ \citenamefont
  {Chattopadhyay}(2020)}]{rajak2020short}%
  \BibitemOpen
  \bibfield  {author} {\bibinfo {author} {\bibfnamefont {R.}~\bibnamefont
  {Rajak}}\ and\ \bibinfo {author} {\bibfnamefont {A.}~\bibnamefont
  {Chattopadhyay}},\ }\href@noop {} {\bibfield  {journal} {\bibinfo  {journal}
  {International journal of environmental health research}\ }\textbf {\bibinfo
  {volume} {30}},\ \bibinfo {pages} {593} (\bibinfo {year} {2020})}\BibitemShut
  {NoStop}%
\bibitem [{\citenamefont {Singh}\ \emph {et~al.}(2016)\citenamefont {Singh},
  \citenamefont {Kumar}, \citenamefont {Kumar}, \citenamefont {Singh},
  \citenamefont {Mina}, \citenamefont {Singh},\ and\ \citenamefont
  {Jain}}]{singh2016}%
  \BibitemOpen
  \bibfield  {author} {\bibinfo {author} {\bibfnamefont {D.}~\bibnamefont
  {Singh}}, \bibinfo {author} {\bibfnamefont {A.}~\bibnamefont {Kumar}},
  \bibinfo {author} {\bibfnamefont {K.}~\bibnamefont {Kumar}}, \bibinfo
  {author} {\bibfnamefont {B.}~\bibnamefont {Singh}}, \bibinfo {author}
  {\bibfnamefont {U.}~\bibnamefont {Mina}}, \bibinfo {author} {\bibfnamefont
  {B.~B.}\ \bibnamefont {Singh}}, \ and\ \bibinfo {author} {\bibfnamefont
  {V.~K.}\ \bibnamefont {Jain}},\ }\href@noop {} {\bibfield  {journal}
  {\bibinfo  {journal} {Science of the Total Environment}\ }\textbf {\bibinfo
  {volume} {572}},\ \bibinfo {pages} {586} (\bibinfo {year}
  {2016})}\BibitemShut {NoStop}%
\bibitem [{\citenamefont {Jayaraman}\ \emph {et~al.}(2007)\citenamefont
  {Jayaraman} \emph {et~al.}}]{jayaraman2007}%
  \BibitemOpen
  \bibfield  {author} {\bibinfo {author} {\bibfnamefont {G.}~\bibnamefont
  {Jayaraman}} \emph {et~al.},\ }\href@noop {} {\bibfield  {journal} {\bibinfo
  {journal} {Environmental monitoring and assessment}\ }\textbf {\bibinfo
  {volume} {135}},\ \bibinfo {pages} {313} (\bibinfo {year}
  {2007})}\BibitemShut {NoStop}%
\bibitem [{\citenamefont {Ware}\ \emph {et~al.}(2016)\citenamefont {Ware},
  \citenamefont {Zhao}, \citenamefont {Koyama}, \citenamefont {May},
  \citenamefont {Matthay}, \citenamefont {Lurmann}, \citenamefont {Balmes},\
  and\ \citenamefont {Calfee}}]{ware2016}%
  \BibitemOpen
  \bibfield  {author} {\bibinfo {author} {\bibfnamefont {L.~B.}\ \bibnamefont
  {Ware}}, \bibinfo {author} {\bibfnamefont {Z.}~\bibnamefont {Zhao}}, \bibinfo
  {author} {\bibfnamefont {T.}~\bibnamefont {Koyama}}, \bibinfo {author}
  {\bibfnamefont {A.~K.}\ \bibnamefont {May}}, \bibinfo {author} {\bibfnamefont
  {M.~A.}\ \bibnamefont {Matthay}}, \bibinfo {author} {\bibfnamefont {F.~W.}\
  \bibnamefont {Lurmann}}, \bibinfo {author} {\bibfnamefont {J.~R.}\
  \bibnamefont {Balmes}}, \ and\ \bibinfo {author} {\bibfnamefont {C.~S.}\
  \bibnamefont {Calfee}},\ }\href@noop {} {\bibfield  {journal} {\bibinfo
  {journal} {American journal of respiratory and critical care medicine}\
  }\textbf {\bibinfo {volume} {193}},\ \bibinfo {pages} {1143} (\bibinfo {year}
  {2016})}\BibitemShut {NoStop}%
\bibitem [{\citenamefont {Conibear}\ \emph {et~al.}(2018)\citenamefont
  {Conibear}, \citenamefont {Butt}, \citenamefont {Knote}, \citenamefont
  {Spracklen},\ and\ \citenamefont {Arnold}}]{conibear2018}%
  \BibitemOpen
  \bibfield  {author} {\bibinfo {author} {\bibfnamefont {L.}~\bibnamefont
  {Conibear}}, \bibinfo {author} {\bibfnamefont {E.~W.}\ \bibnamefont {Butt}},
  \bibinfo {author} {\bibfnamefont {C.}~\bibnamefont {Knote}}, \bibinfo
  {author} {\bibfnamefont {D.~V.}\ \bibnamefont {Spracklen}}, \ and\ \bibinfo
  {author} {\bibfnamefont {S.~R.}\ \bibnamefont {Arnold}},\ }\href@noop {}
  {\bibfield  {journal} {\bibinfo  {journal} {GeoHealth}\ }\textbf {\bibinfo
  {volume} {2}},\ \bibinfo {pages} {334} (\bibinfo {year} {2018})}\BibitemShut
  {NoStop}%
\bibitem [{\citenamefont {Pandey}\ \emph {et~al.}(2021)\citenamefont {Pandey},
  \citenamefont {Brauer}, \citenamefont {Cropper}, \citenamefont
  {Balakrishnan}, \citenamefont {Mathur}, \citenamefont {Dey}, \citenamefont
  {Turkgulu}, \citenamefont {Kumar}, \citenamefont {Khare}, \citenamefont
  {Beig} \emph {et~al.}}]{pandey2021}%
  \BibitemOpen
  \bibfield  {author} {\bibinfo {author} {\bibfnamefont {A.}~\bibnamefont
  {Pandey}}, \bibinfo {author} {\bibfnamefont {M.}~\bibnamefont {Brauer}},
  \bibinfo {author} {\bibfnamefont {M.~L.}\ \bibnamefont {Cropper}}, \bibinfo
  {author} {\bibfnamefont {K.}~\bibnamefont {Balakrishnan}}, \bibinfo {author}
  {\bibfnamefont {P.}~\bibnamefont {Mathur}}, \bibinfo {author} {\bibfnamefont
  {S.}~\bibnamefont {Dey}}, \bibinfo {author} {\bibfnamefont {B.}~\bibnamefont
  {Turkgulu}}, \bibinfo {author} {\bibfnamefont {G.~A.}\ \bibnamefont {Kumar}},
  \bibinfo {author} {\bibfnamefont {M.}~\bibnamefont {Khare}}, \bibinfo
  {author} {\bibfnamefont {G.}~\bibnamefont {Beig}},  \emph {et~al.},\
  }\href@noop {} {\bibfield  {journal} {\bibinfo  {journal} {The Lancet
  Planetary Health}\ }\textbf {\bibinfo {volume} {5}},\ \bibinfo {pages} {e25}
  (\bibinfo {year} {2021})}\BibitemShut {NoStop}%
\bibitem [{\citenamefont {Oksanen}\ \emph {et~al.}(2013)\citenamefont
  {Oksanen}, \citenamefont {Pandey}, \citenamefont {Pandey}, \citenamefont
  {Keski-Saari}, \citenamefont {Kontunen-Soppela},\ and\ \citenamefont
  {Sharma}}]{oksanen13}%
  \BibitemOpen
  \bibfield  {author} {\bibinfo {author} {\bibfnamefont {E.}~\bibnamefont
  {Oksanen}}, \bibinfo {author} {\bibfnamefont {V.}~\bibnamefont {Pandey}},
  \bibinfo {author} {\bibfnamefont {A.~K.}\ \bibnamefont {Pandey}}, \bibinfo
  {author} {\bibfnamefont {S.}~\bibnamefont {Keski-Saari}}, \bibinfo {author}
  {\bibfnamefont {S.}~\bibnamefont {Kontunen-Soppela}}, \ and\ \bibinfo
  {author} {\bibfnamefont {C.}~\bibnamefont {Sharma}},\ }\href {\doibase
  10.1016/j.envpol.2013.02.010} {\bibfield  {journal} {\bibinfo  {journal}
  {ENVIRONMENTAL POLLUTION}\ }\textbf {\bibinfo {volume} {177}},\ \bibinfo
  {pages} {189} (\bibinfo {year} {2013})}\BibitemShut {NoStop}%
\bibitem [{\citenamefont {Van~Dingenen}\ \emph {et~al.}(2009)\citenamefont
  {Van~Dingenen}, \citenamefont {Dentener}, \citenamefont {Raes}, \citenamefont
  {Krol}, \citenamefont {Emberson},\ and\ \citenamefont {Cofala}}]{van09}%
  \BibitemOpen
  \bibfield  {author} {\bibinfo {author} {\bibfnamefont {R.}~\bibnamefont
  {Van~Dingenen}}, \bibinfo {author} {\bibfnamefont {F.~J.}\ \bibnamefont
  {Dentener}}, \bibinfo {author} {\bibfnamefont {F.}~\bibnamefont {Raes}},
  \bibinfo {author} {\bibfnamefont {M.~C.}\ \bibnamefont {Krol}}, \bibinfo
  {author} {\bibfnamefont {L.}~\bibnamefont {Emberson}}, \ and\ \bibinfo
  {author} {\bibfnamefont {J.}~\bibnamefont {Cofala}},\ }\href {\doibase
  10.1016/j.atmosenv.2008.10.033} {\bibfield  {journal} {\bibinfo  {journal}
  {ATMOSPHERIC ENVIRONMENT}\ }\textbf {\bibinfo {volume} {43}},\ \bibinfo
  {pages} {604} (\bibinfo {year} {2009})}\BibitemShut {NoStop}%
\bibitem [{\citenamefont {Yang}\ \emph {et~al.}(2005)\citenamefont {Yang},
  \citenamefont {Cunnold}, \citenamefont {Newchurch},\ and\ \citenamefont
  {Salawitch}}]{yang05}%
  \BibitemOpen
  \bibfield  {author} {\bibinfo {author} {\bibfnamefont {E.}~\bibnamefont
  {Yang}}, \bibinfo {author} {\bibfnamefont {D.}~\bibnamefont {Cunnold}},
  \bibinfo {author} {\bibfnamefont {M.}~\bibnamefont {Newchurch}}, \ and\
  \bibinfo {author} {\bibfnamefont {R.}~\bibnamefont {Salawitch}},\ }\href
  {\doibase 10.1029/2004GL022296} {\bibfield  {journal} {\bibinfo  {journal}
  {GEOPHYSICAL RESEARCH LETTERS}\ }\textbf {\bibinfo {volume} {32}} (\bibinfo
  {year} {2005}),\ 10.1029/2004GL022296}\BibitemShut {NoStop}%
\bibitem [{\citenamefont {Mazzeo}\ \emph {et~al.}(2005)\citenamefont {Mazzeo},
  \citenamefont {Venegas},\ and\ \citenamefont {Choren}}]{mazzeo05}%
  \BibitemOpen
  \bibfield  {author} {\bibinfo {author} {\bibfnamefont {N.}~\bibnamefont
  {Mazzeo}}, \bibinfo {author} {\bibfnamefont {L.}~\bibnamefont {Venegas}}, \
  and\ \bibinfo {author} {\bibfnamefont {H.}~\bibnamefont {Choren}},\ }\href
  {\doibase 10.1016/j.atmosenv.2005.01.029} {\bibfield  {journal} {\bibinfo
  {journal} {ATMOSPHERIC ENVIRONMENT}\ }\textbf {\bibinfo {volume} {39}},\
  \bibinfo {pages} {3055} (\bibinfo {year} {2005})}\BibitemShut {NoStop}%
\bibitem [{\citenamefont {Itano}\ \emph {et~al.}(2007)\citenamefont {Itano},
  \citenamefont {Bandow}, \citenamefont {Takenaka}, \citenamefont {Saitoh},
  \citenamefont {Asayama},\ and\ \citenamefont {Fukuyama}}]{itano07}%
  \BibitemOpen
  \bibfield  {author} {\bibinfo {author} {\bibfnamefont {Y.}~\bibnamefont
  {Itano}}, \bibinfo {author} {\bibfnamefont {H.}~\bibnamefont {Bandow}},
  \bibinfo {author} {\bibfnamefont {N.}~\bibnamefont {Takenaka}}, \bibinfo
  {author} {\bibfnamefont {Y.}~\bibnamefont {Saitoh}}, \bibinfo {author}
  {\bibfnamefont {A.}~\bibnamefont {Asayama}}, \ and\ \bibinfo {author}
  {\bibfnamefont {J.}~\bibnamefont {Fukuyama}},\ }\href {\doibase
  10.1016/j.scitotenv.2007.01.079} {\bibfield  {journal} {\bibinfo  {journal}
  {SCIENCE OF THE TOTAL ENVIRONMENT}\ }\textbf {\bibinfo {volume} {379}},\
  \bibinfo {pages} {46} (\bibinfo {year} {2007})}\BibitemShut {NoStop}%
\bibitem [{\citenamefont {Pudasainee}\ \emph {et~al.}(2006)\citenamefont
  {Pudasainee}, \citenamefont {Sapkota}, \citenamefont {Shrestha},
  \citenamefont {Kaga}, \citenamefont {Kondo},\ and\ \citenamefont
  {Inoue}}]{pudasainee06}%
  \BibitemOpen
  \bibfield  {author} {\bibinfo {author} {\bibfnamefont {D.}~\bibnamefont
  {Pudasainee}}, \bibinfo {author} {\bibfnamefont {B.}~\bibnamefont {Sapkota}},
  \bibinfo {author} {\bibfnamefont {M.~L.}\ \bibnamefont {Shrestha}}, \bibinfo
  {author} {\bibfnamefont {A.}~\bibnamefont {Kaga}}, \bibinfo {author}
  {\bibfnamefont {A.}~\bibnamefont {Kondo}}, \ and\ \bibinfo {author}
  {\bibfnamefont {Y.}~\bibnamefont {Inoue}},\ }\href {\doibase
  10.1016/j.atmosenv.2006.07.011} {\bibfield  {journal} {\bibinfo  {journal}
  {ATMOSPHERIC ENVIRONMENT}\ }\textbf {\bibinfo {volume} {40}},\ \bibinfo
  {pages} {8081} (\bibinfo {year} {2006})}\BibitemShut {NoStop}%
\bibitem [{\citenamefont {Gasmi}\ \emph {et~al.}(2017)\citenamefont {Gasmi},
  \citenamefont {Aljalal}, \citenamefont {Al-Basheer},\ and\ \citenamefont
  {Abdulahi}}]{gasmi2017}%
  \BibitemOpen
  \bibfield  {author} {\bibinfo {author} {\bibfnamefont {K.}~\bibnamefont
  {Gasmi}}, \bibinfo {author} {\bibfnamefont {A.}~\bibnamefont {Aljalal}},
  \bibinfo {author} {\bibfnamefont {W.}~\bibnamefont {Al-Basheer}}, \ and\
  \bibinfo {author} {\bibfnamefont {M.}~\bibnamefont {Abdulahi}},\ }\href@noop
  {} {\bibfield  {journal} {\bibinfo  {journal} {Urban Climate}\ }\textbf
  {\bibinfo {volume} {21}},\ \bibinfo {pages} {232} (\bibinfo {year}
  {2017})}\BibitemShut {NoStop}%
\bibitem [{\citenamefont {Tiwari}\ \emph {et~al.}(2015)\citenamefont {Tiwari},
  \citenamefont {Dahiya},\ and\ \citenamefont {Kumar}}]{tiwari2015}%
  \BibitemOpen
  \bibfield  {author} {\bibinfo {author} {\bibfnamefont {S.}~\bibnamefont
  {Tiwari}}, \bibinfo {author} {\bibfnamefont {A.}~\bibnamefont {Dahiya}}, \
  and\ \bibinfo {author} {\bibfnamefont {N.}~\bibnamefont {Kumar}},\
  }\href@noop {} {\bibfield  {journal} {\bibinfo  {journal} {Atmospheric
  Research}\ }\textbf {\bibinfo {volume} {157}},\ \bibinfo {pages} {119}
  (\bibinfo {year} {2015})}\BibitemShut {NoStop}%
\bibitem [{\citenamefont {Sillman}(1999)}]{sillman99}%
  \BibitemOpen
  \bibfield  {author} {\bibinfo {author} {\bibfnamefont {S.}~\bibnamefont
  {Sillman}},\ }\href {\doibase 10.1016/S1352-2310(98)00345-8} {\bibfield
  {journal} {\bibinfo  {journal} {ATMOSPHERIC ENVIRONMENT}\ }\textbf {\bibinfo
  {volume} {33}},\ \bibinfo {pages} {1821} (\bibinfo {year}
  {1999})}\BibitemShut {NoStop}%
\bibitem [{\citenamefont {Londhe}\ \emph {et~al.}(2008)\citenamefont {Londhe},
  \citenamefont {Jadhav}, \citenamefont {Buchunde},\ and\ \citenamefont
  {Kartha}}]{londhe08}%
  \BibitemOpen
  \bibfield  {author} {\bibinfo {author} {\bibfnamefont {A.~L.}\ \bibnamefont
  {Londhe}}, \bibinfo {author} {\bibfnamefont {D.~B.}\ \bibnamefont {Jadhav}},
  \bibinfo {author} {\bibfnamefont {P.~S.}\ \bibnamefont {Buchunde}}, \ and\
  \bibinfo {author} {\bibfnamefont {M.~J.}\ \bibnamefont {Kartha}},\
  }\href@noop {} {\bibfield  {journal} {\bibinfo  {journal} {CURRENT SCIENCE}\
  }\textbf {\bibinfo {volume} {95}},\ \bibinfo {pages} {1724} (\bibinfo {year}
  {2008})}\BibitemShut {NoStop}%
\bibitem [{\citenamefont {Camalier}\ \emph {et~al.}(2007)\citenamefont
  {Camalier}, \citenamefont {Cox},\ and\ \citenamefont
  {Dolwick}}]{CAMALIER20077127}%
  \BibitemOpen
  \bibfield  {author} {\bibinfo {author} {\bibfnamefont {L.}~\bibnamefont
  {Camalier}}, \bibinfo {author} {\bibfnamefont {W.}~\bibnamefont {Cox}}, \
  and\ \bibinfo {author} {\bibfnamefont {P.}~\bibnamefont {Dolwick}},\ }\href
  {\doibase https://doi.org/10.1016/j.atmosenv.2007.04.061} {\bibfield
  {journal} {\bibinfo  {journal} {Atmospheric Environment}\ }\textbf {\bibinfo
  {volume} {41}},\ \bibinfo {pages} {7127} (\bibinfo {year}
  {2007})}\BibitemShut {NoStop}%
\bibitem [{\citenamefont {Li}\ \emph {et~al.}(2021)\citenamefont {Li},
  \citenamefont {Yu}, \citenamefont {Chen}, \citenamefont {Li}, \citenamefont
  {Zhang}, \citenamefont {Wang}, \citenamefont {Liu}, \citenamefont {Li},
  \citenamefont {Lichtfouse}, \citenamefont {Rosenfeld} \emph
  {et~al.}}]{li2021large}%
  \BibitemOpen
  \bibfield  {author} {\bibinfo {author} {\bibfnamefont {M.}~\bibnamefont
  {Li}}, \bibinfo {author} {\bibfnamefont {S.}~\bibnamefont {Yu}}, \bibinfo
  {author} {\bibfnamefont {X.}~\bibnamefont {Chen}}, \bibinfo {author}
  {\bibfnamefont {Z.}~\bibnamefont {Li}}, \bibinfo {author} {\bibfnamefont
  {Y.}~\bibnamefont {Zhang}}, \bibinfo {author} {\bibfnamefont
  {L.}~\bibnamefont {Wang}}, \bibinfo {author} {\bibfnamefont {W.}~\bibnamefont
  {Liu}}, \bibinfo {author} {\bibfnamefont {P.}~\bibnamefont {Li}}, \bibinfo
  {author} {\bibfnamefont {E.}~\bibnamefont {Lichtfouse}}, \bibinfo {author}
  {\bibfnamefont {D.}~\bibnamefont {Rosenfeld}},  \emph {et~al.},\ }\href@noop
  {} {\bibfield  {journal} {\bibinfo  {journal} {Environmental Chemistry
  Letters}\ }\textbf {\bibinfo {volume} {19}},\ \bibinfo {pages} {3981}
  (\bibinfo {year} {2021})}\BibitemShut {NoStop}%
\bibitem [{\citenamefont {Nishanth}\ \emph
  {et~al.}(2012{\natexlab{a}})\citenamefont {Nishanth}, \citenamefont {Kumar},\
  and\ \citenamefont {Valsaraj}}]{nishanth12}%
  \BibitemOpen
  \bibfield  {author} {\bibinfo {author} {\bibfnamefont {T.}~\bibnamefont
  {Nishanth}}, \bibinfo {author} {\bibfnamefont {M.~K.~S.}\ \bibnamefont
  {Kumar}}, \ and\ \bibinfo {author} {\bibfnamefont {K.~T.}\ \bibnamefont
  {Valsaraj}},\ }\href {\doibase 10.1007/s10874-012-9234-5} {\bibfield
  {journal} {\bibinfo  {journal} {JOURNAL OF ATMOSPHERIC CHEMISTRY}\ }\textbf
  {\bibinfo {volume} {69}},\ \bibinfo {pages} {101} (\bibinfo {year}
  {2012}{\natexlab{a}})}\BibitemShut {NoStop}%
\bibitem [{Note3()}]{Note3}%
  \BibitemOpen
  \bibinfo {note} {\protect \url
  {https://app.cpcbccr.com/ccr//caaqm-dashboard-all/caaqm-landing}}\BibitemShut
  {NoStop}%
\bibitem [{Note4()}]{Note4}%
  \BibitemOpen
  \bibinfo {note} {\protect \url
  {https://en.climate-data.org/asia/india/madhya-pradesh/bhopal-2833/}}\BibitemShut
  {NoStop}%
\bibitem [{\citenamefont {Longmore}\ \emph {et~al.}(2019)\citenamefont
  {Longmore}, \citenamefont {Lui}, \citenamefont {Naik}, \citenamefont {Breen},
  \citenamefont {Jalaludin},\ and\ \citenamefont
  {Gargiulo}}]{longmore2019comparison}%
  \BibitemOpen
  \bibfield  {author} {\bibinfo {author} {\bibfnamefont {S.~K.}\ \bibnamefont
  {Longmore}}, \bibinfo {author} {\bibfnamefont {G.~Y.}\ \bibnamefont {Lui}},
  \bibinfo {author} {\bibfnamefont {G.}~\bibnamefont {Naik}}, \bibinfo {author}
  {\bibfnamefont {P.~P.}\ \bibnamefont {Breen}}, \bibinfo {author}
  {\bibfnamefont {B.}~\bibnamefont {Jalaludin}}, \ and\ \bibinfo {author}
  {\bibfnamefont {G.~D.}\ \bibnamefont {Gargiulo}},\ }\href@noop {} {\bibfield
  {journal} {\bibinfo  {journal} {Sensors}\ }\textbf {\bibinfo {volume} {19}},\
  \bibinfo {pages} {1874} (\bibinfo {year} {2019})}\BibitemShut {NoStop}%
\bibitem [{\citenamefont {Zainuri}\ \emph {et~al.}(2015)\citenamefont
  {Zainuri}, \citenamefont {Jemain},\ and\ \citenamefont
  {Muda}}]{zainuri2015comparison}%
  \BibitemOpen
  \bibfield  {author} {\bibinfo {author} {\bibfnamefont {N.~A.}\ \bibnamefont
  {Zainuri}}, \bibinfo {author} {\bibfnamefont {A.~A.}\ \bibnamefont {Jemain}},
  \ and\ \bibinfo {author} {\bibfnamefont {N.}~\bibnamefont {Muda}},\
  }\href@noop {} {\bibfield  {journal} {\bibinfo  {journal} {Sains Malaysiana}\
  }\textbf {\bibinfo {volume} {44}},\ \bibinfo {pages} {449} (\bibinfo {year}
  {2015})}\BibitemShut {NoStop}%
\bibitem [{\citenamefont {Leighton}(1961)}]{Leighton61}%
  \BibitemOpen
  \bibfield  {author} {\bibinfo {author} {\bibfnamefont {P.~A.}\ \bibnamefont
  {Leighton}},\ }\href@noop {} {{\emph {\bibinfo {title}
  {Photochemistry of air pollution}}}},\ Physical chemistry (Academic Press) ;
  v.9\ (\bibinfo  {publisher} {Academic Press},\ \bibinfo {address} {New York ;
  London},\ \bibinfo {year} {1961})\BibitemShut {NoStop}%
\bibitem [{\citenamefont {Leighton}(2012)}]{leighton2012}%
  \BibitemOpen
  \bibfield  {author} {\bibinfo {author} {\bibfnamefont {P.}~\bibnamefont
  {Leighton}},\ }\href@noop {} {\emph {\bibinfo {title} {Photochemistry of air
  pollution}}}\ (\bibinfo  {publisher} {Elsevier},\ \bibinfo {year}
  {2012})\BibitemShut {NoStop}%
\bibitem [{\citenamefont {Seinfeld}\ and\ \citenamefont
  {Pandis}(1998)}]{seinfeld98}%
  \BibitemOpen
  \bibfield  {author} {\bibinfo {author} {\bibfnamefont {J.~H.}\ \bibnamefont
  {Seinfeld}}\ and\ \bibinfo {author} {\bibfnamefont {S.~N.}\ \bibnamefont
  {Pandis}},\ }\href@noop {} {\bibfield  {journal} {\bibinfo  {journal}
  {Atmospheric chemistry and physics}\ }\textbf {\bibinfo {volume} {1326}}
  (\bibinfo {year} {1998})}\BibitemShut {NoStop}%
\bibitem [{\citenamefont {Latini}\ \emph {et~al.}(2002)\citenamefont {Latini},
  \citenamefont {Grifoni},\ and\ \citenamefont
  {Passerini}}]{latini2002influence}%
  \BibitemOpen
  \bibfield  {author} {\bibinfo {author} {\bibfnamefont {G.}~\bibnamefont
  {Latini}}, \bibinfo {author} {\bibfnamefont {R.~C.}\ \bibnamefont {Grifoni}},
  \ and\ \bibinfo {author} {\bibfnamefont {G.}~\bibnamefont {Passerini}},\
  }\href@noop {} {\bibfield  {journal} {\bibinfo  {journal} {WIT Transactions
  on Ecology and the Environment}\ }\textbf {\bibinfo {volume} {53}} (\bibinfo
  {year} {2002})}\BibitemShut {NoStop}%
\bibitem [{\citenamefont {Lee~Rodgers}\ and\ \citenamefont
  {Nicewander}(1988)}]{lee1988thirteen}%
  \BibitemOpen
  \bibfield  {author} {\bibinfo {author} {\bibfnamefont {J.}~\bibnamefont
  {Lee~Rodgers}}\ and\ \bibinfo {author} {\bibfnamefont {W.~A.}\ \bibnamefont
  {Nicewander}},\ }\href@noop {} {\bibfield  {journal} {\bibinfo  {journal}
  {The American Statistician}\ }\textbf {\bibinfo {volume} {42}},\ \bibinfo
  {pages} {59} (\bibinfo {year} {1988})}\BibitemShut {NoStop}%
\bibitem [{\citenamefont {Clapp}\ and\ \citenamefont {Jenkin}(2001)}]{clapp01}%
  \BibitemOpen
  \bibfield  {author} {\bibinfo {author} {\bibfnamefont {L.}~\bibnamefont
  {Clapp}}\ and\ \bibinfo {author} {\bibfnamefont {M.}~\bibnamefont {Jenkin}},\
  }\href {\doibase 10.1016/S1352-2310(01)00378-8} {\bibfield  {journal}
  {\bibinfo  {journal} {ATMOSPHERIC ENVIRONMENT}\ }\textbf {\bibinfo {volume}
  {35}},\ \bibinfo {pages} {6391} (\bibinfo {year} {2001})}\BibitemShut
  {NoStop}%
\bibitem [{\citenamefont {Kley}\ \emph {et~al.}(1994)\citenamefont {Kley},
  \citenamefont {Geiss},\ and\ \citenamefont {Mohnen}}]{kley1994}%
  \BibitemOpen
  \bibfield  {author} {\bibinfo {author} {\bibfnamefont {D.}~\bibnamefont
  {Kley}}, \bibinfo {author} {\bibfnamefont {H.}~\bibnamefont {Geiss}}, \ and\
  \bibinfo {author} {\bibfnamefont {V.~A.}\ \bibnamefont {Mohnen}},\
  }\href@noop {} {\bibfield  {journal} {\bibinfo  {journal} {Atmospheric
  Environment}\ }\textbf {\bibinfo {volume} {28}},\ \bibinfo {pages} {149}
  (\bibinfo {year} {1994})}\BibitemShut {NoStop}%
\bibitem [{\citenamefont {Tiwari}\ \emph {et~al.}(2014)\citenamefont {Tiwari},
  \citenamefont {Srivastava}, \citenamefont {Chate}, \citenamefont {Safai},
  \citenamefont {Bisht}, \citenamefont {Srivastava},\ and\ \citenamefont
  {Beig}}]{tiwari14}%
  \BibitemOpen
  \bibfield  {author} {\bibinfo {author} {\bibfnamefont {S.}~\bibnamefont
  {Tiwari}}, \bibinfo {author} {\bibfnamefont {A.~K.}\ \bibnamefont
  {Srivastava}}, \bibinfo {author} {\bibfnamefont {D.~M.}\ \bibnamefont
  {Chate}}, \bibinfo {author} {\bibfnamefont {P.~D.}\ \bibnamefont {Safai}},
  \bibinfo {author} {\bibfnamefont {D.~S.}\ \bibnamefont {Bisht}}, \bibinfo
  {author} {\bibfnamefont {M.~K.}\ \bibnamefont {Srivastava}}, \ and\ \bibinfo
  {author} {\bibfnamefont {G.}~\bibnamefont {Beig}},\ }\href {\doibase
  10.1016/j.atmosenv.2014.03.064} {\bibfield  {journal} {\bibinfo  {journal}
  {ATMOSPHERIC ENVIRONMENT}\ }\textbf {\bibinfo {volume} {92}},\ \bibinfo
  {pages} {60} (\bibinfo {year} {2014})}\BibitemShut {NoStop}%
\bibitem [{\citenamefont {Han}\ \emph {et~al.}(2011)\citenamefont {Han},
  \citenamefont {Bian}, \citenamefont {Feng}, \citenamefont {Liu},
  \citenamefont {Li}, \citenamefont {Zeng},\ and\ \citenamefont
  {Zhang}}]{han11}%
  \BibitemOpen
  \bibfield  {author} {\bibinfo {author} {\bibfnamefont {S.}~\bibnamefont
  {Han}}, \bibinfo {author} {\bibfnamefont {H.}~\bibnamefont {Bian}}, \bibinfo
  {author} {\bibfnamefont {Y.}~\bibnamefont {Feng}}, \bibinfo {author}
  {\bibfnamefont {A.}~\bibnamefont {Liu}}, \bibinfo {author} {\bibfnamefont
  {X.}~\bibnamefont {Li}}, \bibinfo {author} {\bibfnamefont {F.}~\bibnamefont
  {Zeng}}, \ and\ \bibinfo {author} {\bibfnamefont {X.}~\bibnamefont {Zhang}},\
  }\href {\doibase 10.4209/aaqr.2010.07.0055} {\bibfield  {journal} {\bibinfo
  {journal} {AEROSOL AND AIR QUALITY RESEARCH}\ }\textbf {\bibinfo {volume}
  {11}},\ \bibinfo {pages} {128} (\bibinfo {year} {2011})}\BibitemShut
  {NoStop}%
\bibitem [{\citenamefont {Hassan}\ \emph {et~al.}(2013)\citenamefont {Hassan},
  \citenamefont {Basahi}, \citenamefont {Ismail},\ and\ \citenamefont
  {Habeebullah}}]{hassan13}%
  \BibitemOpen
  \bibfield  {author} {\bibinfo {author} {\bibfnamefont {I.~A.}\ \bibnamefont
  {Hassan}}, \bibinfo {author} {\bibfnamefont {J.~M.}\ \bibnamefont {Basahi}},
  \bibinfo {author} {\bibfnamefont {I.~M.}\ \bibnamefont {Ismail}}, \ and\
  \bibinfo {author} {\bibfnamefont {T.~M.}\ \bibnamefont {Habeebullah}},\
  }\href {\doibase 10.4209/aaqr.2013.01.0007} {\bibfield  {journal} {\bibinfo
  {journal} {AEROSOL AND AIR QUALITY RESEARCH}\ }\textbf {\bibinfo {volume}
  {13}},\ \bibinfo {pages} {1712} (\bibinfo {year} {2013})}\BibitemShut
  {NoStop}%
\bibitem [{\citenamefont {Nagpure}\ \emph {et~al.}(2013)\citenamefont
  {Nagpure}, \citenamefont {Sharma},\ and\ \citenamefont {Gurjar}}]{nagpure13}%
  \BibitemOpen
  \bibfield  {author} {\bibinfo {author} {\bibfnamefont {A.~S.}\ \bibnamefont
  {Nagpure}}, \bibinfo {author} {\bibfnamefont {K.}~\bibnamefont {Sharma}}, \
  and\ \bibinfo {author} {\bibfnamefont {B.~R.}\ \bibnamefont {Gurjar}},\
  }\href {\doibase 10.1016/j.uclim.2013.04.005} {\bibfield  {journal} {\bibinfo
   {journal} {URBAN CLIMATE}\ }\textbf {\bibinfo {volume} {4}},\ \bibinfo
  {pages} {61} (\bibinfo {year} {2013})}\BibitemShut {NoStop}%
\bibitem [{\citenamefont {Reddy}\ \emph {et~al.}(2012)\citenamefont {Reddy},
  \citenamefont {Kumar}, \citenamefont {Balakrishnaiah}, \citenamefont {Gopal},
  \citenamefont {Reddy}, \citenamefont {Sivakumar}, \citenamefont {Lingaswamy},
  \citenamefont {Arafath}, \citenamefont {Umadevi}, \citenamefont {Kumari},
  \citenamefont {Ahammed},\ and\ \citenamefont {Lal}}]{reddy12}%
  \BibitemOpen
  \bibfield  {author} {\bibinfo {author} {\bibfnamefont {B.~S.~K.}\
  \bibnamefont {Reddy}}, \bibinfo {author} {\bibfnamefont {K.~R.}\ \bibnamefont
  {Kumar}}, \bibinfo {author} {\bibfnamefont {G.}~\bibnamefont
  {Balakrishnaiah}}, \bibinfo {author} {\bibfnamefont {K.~R.}\ \bibnamefont
  {Gopal}}, \bibinfo {author} {\bibfnamefont {R.~R.}\ \bibnamefont {Reddy}},
  \bibinfo {author} {\bibfnamefont {V.}~\bibnamefont {Sivakumar}}, \bibinfo
  {author} {\bibfnamefont {A.~P.}\ \bibnamefont {Lingaswamy}}, \bibinfo
  {author} {\bibfnamefont {S.~M.}\ \bibnamefont {Arafath}}, \bibinfo {author}
  {\bibfnamefont {K.}~\bibnamefont {Umadevi}}, \bibinfo {author} {\bibfnamefont
  {S.~P.}\ \bibnamefont {Kumari}}, \bibinfo {author} {\bibfnamefont {Y.~N.}\
  \bibnamefont {Ahammed}}, \ and\ \bibinfo {author} {\bibfnamefont
  {S.}~\bibnamefont {Lal}},\ }\href {\doibase 10.4209/aaqr.2012.03.0055}
  {\bibfield  {journal} {\bibinfo  {journal} {AEROSOL AND AIR QUALITY
  RESEARCH}\ }\textbf {\bibinfo {volume} {12}},\ \bibinfo {pages} {1081}
  (\bibinfo {year} {2012})}\BibitemShut {NoStop}%
\bibitem [{\citenamefont {Badarinath}\ \emph {et~al.}(2007)\citenamefont
  {Badarinath}, \citenamefont {Kharol}, \citenamefont {Chand}, \citenamefont
  {Parvathi}, \citenamefont {Anasuya},\ and\ \citenamefont
  {Jyothsna}}]{badarinath07}%
  \BibitemOpen
  \bibfield  {author} {\bibinfo {author} {\bibfnamefont {K.~V.~S.}\
  \bibnamefont {Badarinath}}, \bibinfo {author} {\bibfnamefont {S.~K.}\
  \bibnamefont {Kharol}}, \bibinfo {author} {\bibfnamefont {T.~R.~K.}\
  \bibnamefont {Chand}}, \bibinfo {author} {\bibfnamefont {Y.~G.}\ \bibnamefont
  {Parvathi}}, \bibinfo {author} {\bibfnamefont {T.}~\bibnamefont {Anasuya}}, \
  and\ \bibinfo {author} {\bibfnamefont {A.~N.}\ \bibnamefont {Jyothsna}},\
  }\href {\doibase 10.1016/j.atmosres.2006.10.004} {\bibfield  {journal}
  {\bibinfo  {journal} {ATMOSPHERIC RESEARCH}\ }\textbf {\bibinfo {volume}
  {85}},\ \bibinfo {pages} {18} (\bibinfo {year} {2007})}\BibitemShut {NoStop}%
\bibitem [{\citenamefont {He}\ and\ \citenamefont {Lu}(2012)}]{he12}%
  \BibitemOpen
  \bibfield  {author} {\bibinfo {author} {\bibfnamefont {H.-d.}\ \bibnamefont
  {He}}\ and\ \bibinfo {author} {\bibfnamefont {W.-Z.}\ \bibnamefont {Lu}},\
  }\href {\doibase 10.1016/j.buildenv.2011.09.019} {\bibfield  {journal}
  {\bibinfo  {journal} {BUILDING AND ENVIRONMENT}\ }\textbf {\bibinfo {volume}
  {49}},\ \bibinfo {pages} {97} (\bibinfo {year} {2012})}\BibitemShut {NoStop}%
\bibitem [{\citenamefont {Jacob}\ and\ \citenamefont {Winner}(2009)}]{jacob09}%
  \BibitemOpen
  \bibfield  {author} {\bibinfo {author} {\bibfnamefont {D.~J.}\ \bibnamefont
  {Jacob}}\ and\ \bibinfo {author} {\bibfnamefont {D.~A.}\ \bibnamefont
  {Winner}},\ }\href {\doibase 10.1016/j.atmosenv.2008.09.051} {\bibfield
  {journal} {\bibinfo  {journal} {ATMOSPHERIC ENVIRONMENT}\ }\textbf {\bibinfo
  {volume} {43}},\ \bibinfo {pages} {51} (\bibinfo {year} {2009})}\BibitemShut
  {NoStop}%
\bibitem [{\citenamefont {Teixeira}\ \emph {et~al.}(2009)\citenamefont
  {Teixeira}, \citenamefont {de~Santana}, \citenamefont {Wiegand},\ and\
  \citenamefont {Fachel}}]{teixeira09}%
  \BibitemOpen
  \bibfield  {author} {\bibinfo {author} {\bibfnamefont {E.~C.}\ \bibnamefont
  {Teixeira}}, \bibinfo {author} {\bibfnamefont {E.~R.}\ \bibnamefont
  {de~Santana}}, \bibinfo {author} {\bibfnamefont {F.}~\bibnamefont {Wiegand}},
  \ and\ \bibinfo {author} {\bibfnamefont {J.}~\bibnamefont {Fachel}},\ }\href
  {\doibase 10.1016/j.atmosenv.2008.12.051} {\bibfield  {journal} {\bibinfo
  {journal} {ATMOSPHERIC ENVIRONMENT}\ }\textbf {\bibinfo {volume} {43}},\
  \bibinfo {pages} {2213} (\bibinfo {year} {2009})}\BibitemShut {NoStop}%
\bibitem [{\citenamefont {Gaur}\ \emph {et~al.}(2014)\citenamefont {Gaur},
  \citenamefont {Tripathi}, \citenamefont {Kanawade}, \citenamefont {Tare},\
  and\ \citenamefont {Shukla}}]{gaur2014}%
  \BibitemOpen
  \bibfield  {author} {\bibinfo {author} {\bibfnamefont {A.}~\bibnamefont
  {Gaur}}, \bibinfo {author} {\bibfnamefont {S.}~\bibnamefont {Tripathi}},
  \bibinfo {author} {\bibfnamefont {V.}~\bibnamefont {Kanawade}}, \bibinfo
  {author} {\bibfnamefont {V.}~\bibnamefont {Tare}}, \ and\ \bibinfo {author}
  {\bibfnamefont {S.}~\bibnamefont {Shukla}},\ }\href@noop {} {\bibfield
  {journal} {\bibinfo  {journal} {Journal of Atmospheric Chemistry}\ }\textbf
  {\bibinfo {volume} {71}},\ \bibinfo {pages} {283} (\bibinfo {year}
  {2014})}\BibitemShut {NoStop}%
\bibitem [{\citenamefont {Khoder}(2009)}]{khoder2009}%
  \BibitemOpen
  \bibfield  {author} {\bibinfo {author} {\bibfnamefont {M.~I.}\ \bibnamefont
  {Khoder}},\ }\href@noop {} {\bibfield  {journal} {\bibinfo  {journal}
  {Environmental Monitoring and Assessment}\ }\textbf {\bibinfo {volume}
  {149}},\ \bibinfo {pages} {349} (\bibinfo {year} {2009})}\BibitemShut
  {NoStop}%
\bibitem [{\citenamefont {Tian}\ \emph {et~al.}(2020)\citenamefont {Tian},
  \citenamefont {Fan}, \citenamefont {Jin}, \citenamefont {Mao}, \citenamefont
  {Geng}, \citenamefont {Hou}, \citenamefont {Zhang},\ and\ \citenamefont
  {Zhang}}]{tian2020}%
  \BibitemOpen
  \bibfield  {author} {\bibinfo {author} {\bibfnamefont {D.}~\bibnamefont
  {Tian}}, \bibinfo {author} {\bibfnamefont {J.}~\bibnamefont {Fan}}, \bibinfo
  {author} {\bibfnamefont {H.}~\bibnamefont {Jin}}, \bibinfo {author}
  {\bibfnamefont {H.}~\bibnamefont {Mao}}, \bibinfo {author} {\bibfnamefont
  {D.}~\bibnamefont {Geng}}, \bibinfo {author} {\bibfnamefont {S.}~\bibnamefont
  {Hou}}, \bibinfo {author} {\bibfnamefont {P.}~\bibnamefont {Zhang}}, \ and\
  \bibinfo {author} {\bibfnamefont {Y.}~\bibnamefont {Zhang}},\ }\href@noop {}
  {\bibfield  {journal} {\bibinfo  {journal} {Journal of Geophysical Research:
  Atmospheres}\ }\textbf {\bibinfo {volume} {125}},\ \bibinfo {pages}
  {e2019JD031931} (\bibinfo {year} {2020})}\BibitemShut {NoStop}%
\bibitem [{\citenamefont {Nishanth}\ \emph
  {et~al.}(2012{\natexlab{b}})\citenamefont {Nishanth}, \citenamefont
  {Satheesh~Kumar},\ and\ \citenamefont {Valsaraj}}]{nishanth2012}%
  \BibitemOpen
  \bibfield  {author} {\bibinfo {author} {\bibfnamefont {T.}~\bibnamefont
  {Nishanth}}, \bibinfo {author} {\bibfnamefont {M.}~\bibnamefont
  {Satheesh~Kumar}}, \ and\ \bibinfo {author} {\bibfnamefont {K.}~\bibnamefont
  {Valsaraj}},\ }\href@noop {} {\bibfield  {journal} {\bibinfo  {journal}
  {Journal of Atmospheric Chemistry}\ }\textbf {\bibinfo {volume} {69}},\
  \bibinfo {pages} {101} (\bibinfo {year} {2012}{\natexlab{b}})}\BibitemShut
  {NoStop}%
\bibitem [{\citenamefont {Swamy}\ \emph {et~al.}(2012)\citenamefont {Swamy},
  \citenamefont {Venkanna}, \citenamefont {Nikhil}, \citenamefont {Chitanya},
  \citenamefont {Sinha}, \citenamefont {Ramakrishna},\ and\ \citenamefont
  {Rao}}]{swamy12}%
  \BibitemOpen
  \bibfield  {author} {\bibinfo {author} {\bibfnamefont {Y.~V.}\ \bibnamefont
  {Swamy}}, \bibinfo {author} {\bibfnamefont {R.}~\bibnamefont {Venkanna}},
  \bibinfo {author} {\bibfnamefont {G.~N.}\ \bibnamefont {Nikhil}}, \bibinfo
  {author} {\bibfnamefont {D.~N. S.~K.}\ \bibnamefont {Chitanya}}, \bibinfo
  {author} {\bibfnamefont {P.~R.}\ \bibnamefont {Sinha}}, \bibinfo {author}
  {\bibfnamefont {M.}~\bibnamefont {Ramakrishna}}, \ and\ \bibinfo {author}
  {\bibfnamefont {A.~G.}\ \bibnamefont {Rao}},\ }\href {\doibase
  10.4209/aaqr.2012.01.0019} {\bibfield  {journal} {\bibinfo  {journal}
  {AEROSOL AND AIR QUALITY RESEARCH}\ }\textbf {\bibinfo {volume} {12}},\
  \bibinfo {pages} {662} (\bibinfo {year} {2012})}\BibitemShut {NoStop}%
\bibitem [{\citenamefont {Kumar}\ \emph {et~al.}(2015)\citenamefont {Kumar},
  \citenamefont {Singh}, \citenamefont {Singh}, \citenamefont {Singh},
  \citenamefont {Anandam}, \citenamefont {Kumar},\ and\ \citenamefont
  {Jain}}]{kumar2015}%
  \BibitemOpen
  \bibfield  {author} {\bibinfo {author} {\bibfnamefont {A.}~\bibnamefont
  {Kumar}}, \bibinfo {author} {\bibfnamefont {D.}~\bibnamefont {Singh}},
  \bibinfo {author} {\bibfnamefont {B.~P.}\ \bibnamefont {Singh}}, \bibinfo
  {author} {\bibfnamefont {M.}~\bibnamefont {Singh}}, \bibinfo {author}
  {\bibfnamefont {K.}~\bibnamefont {Anandam}}, \bibinfo {author} {\bibfnamefont
  {K.}~\bibnamefont {Kumar}}, \ and\ \bibinfo {author} {\bibfnamefont
  {V.}~\bibnamefont {Jain}},\ }\href@noop {} {\bibfield  {journal} {\bibinfo
  {journal} {Air Quality, Atmosphere \& Health}\ }\textbf {\bibinfo {volume}
  {8}},\ \bibinfo {pages} {391} (\bibinfo {year} {2015})}\BibitemShut {NoStop}%
\bibitem [{\citenamefont {Jones}\ \emph {et~al.}(2010)\citenamefont {Jones},
  \citenamefont {Harrison},\ and\ \citenamefont {Baker}}]{jones10}%
  \BibitemOpen
  \bibfield  {author} {\bibinfo {author} {\bibfnamefont {A.~M.}\ \bibnamefont
  {Jones}}, \bibinfo {author} {\bibfnamefont {R.~M.}\ \bibnamefont {Harrison}},
  \ and\ \bibinfo {author} {\bibfnamefont {J.}~\bibnamefont {Baker}},\ }\href
  {\doibase 10.1016/j.atmosenv.2010.01.007} {\bibfield  {journal} {\bibinfo
  {journal} {ATMOSPHERIC ENVIRONMENT}\ }\textbf {\bibinfo {volume} {44}},\
  \bibinfo {pages} {1682} (\bibinfo {year} {2010})}\BibitemShut {NoStop}%
\bibitem [{\citenamefont {Korhale}\ \emph {et~al.}(2022)\citenamefont
  {Korhale}, \citenamefont {Anand}, \citenamefont {Panicker},\ and\
  \citenamefont {Beig}}]{korhale2022}%
  \BibitemOpen
  \bibfield  {author} {\bibinfo {author} {\bibfnamefont {N.}~\bibnamefont
  {Korhale}}, \bibinfo {author} {\bibfnamefont {V.}~\bibnamefont {Anand}},
  \bibinfo {author} {\bibfnamefont {A.}~\bibnamefont {Panicker}}, \ and\
  \bibinfo {author} {\bibfnamefont {G.}~\bibnamefont {Beig}},\ }\href@noop {}
  {\bibfield  {journal} {\bibinfo  {journal} {International Journal of
  Environmental Science and Technology}\ ,\ \bibinfo {pages} {1}} (\bibinfo
  {year} {2022})}\BibitemShut {NoStop}%
\bibitem [{\citenamefont {Elminir}(2005)}]{elminir05}%
  \BibitemOpen
  \bibfield  {author} {\bibinfo {author} {\bibfnamefont {H.}~\bibnamefont
  {Elminir}},\ }\href {\doibase 10.1016/j.scitotenv.2005.01.043} {\bibfield
  {journal} {\bibinfo  {journal} {SCIENCE OF THE TOTAL ENVIRONMENT}\ }\textbf
  {\bibinfo {volume} {350}},\ \bibinfo {pages} {225} (\bibinfo {year}
  {2005})}\BibitemShut {NoStop}%
\bibitem [{\citenamefont {Mavroidis}\ and\ \citenamefont
  {Ilia}(2012)}]{mavroidis12}%
  \BibitemOpen
  \bibfield  {author} {\bibinfo {author} {\bibfnamefont {I.}~\bibnamefont
  {Mavroidis}}\ and\ \bibinfo {author} {\bibfnamefont {M.}~\bibnamefont
  {Ilia}},\ }\href {\doibase 10.1016/j.atmosenv.2012.09.030} {\bibfield
  {journal} {\bibinfo  {journal} {ATMOSPHERIC ENVIRONMENT}\ }\textbf {\bibinfo
  {volume} {63}},\ \bibinfo {pages} {135} (\bibinfo {year} {2012})}\BibitemShut
  {NoStop}%
\bibitem [{\citenamefont {Ma}\ and\ \citenamefont {van Weele}(2000)}]{ma2000}%
  \BibitemOpen
  \bibfield  {author} {\bibinfo {author} {\bibfnamefont {J.}~\bibnamefont
  {Ma}}\ and\ \bibinfo {author} {\bibfnamefont {M.}~\bibnamefont {van Weele}},\
  }\href@noop {} {\bibfield  {journal} {\bibinfo  {journal} {Chemosphere-Global
  Change Science}\ }\textbf {\bibinfo {volume} {2}},\ \bibinfo {pages} {23}
  (\bibinfo {year} {2000})}\BibitemShut {NoStop}%
\end{thebibliography}%

\section*{Appendix}

\subsection{Time series of diurnal variation $O_3$, $NO$, $NO_2$, and $NO_x$}\label{secA1}
 For diurnal dependency analysis, we normalize the data-set by the maximum concentration of the each pollutants for each day given as the following equation:
\begin{equation}
    x^{(k)}(h)=\frac{1}{365}\sum_{d=1}^{365}\frac{x^{(k)}(d,h)}{max_h[x^{(k)}(d,h)]}\label{eqn1}
\end{equation}
\begin{figure*}{H}
	\centering
	\includegraphics[width=0.8\textwidth]{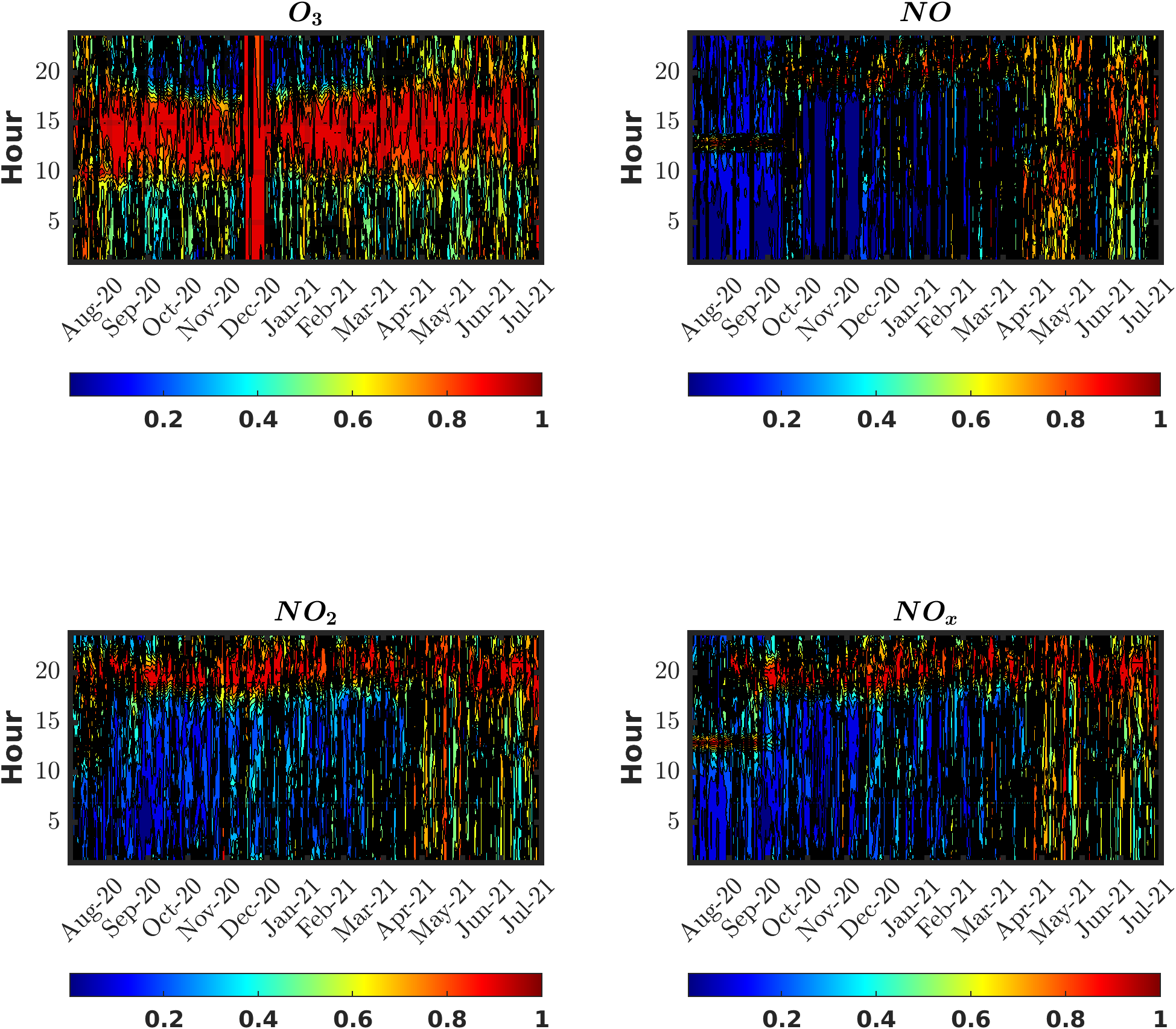}
\caption{Figure shows the time series contour plot of diurnal variation of $O_3$, $NO$, $NO_2$, and $NO_x$ at Commercial for the full time period considered.}\label{fig7}

%The strong diurnal variations in $NO_2$ and PM2.5 are indicative of rush hour related emissions and photochemical production in case of $O_3$.}
\end{figure*}
\begin{figure*}{H}
	\centering
	\includegraphics[width=0.8\textwidth]{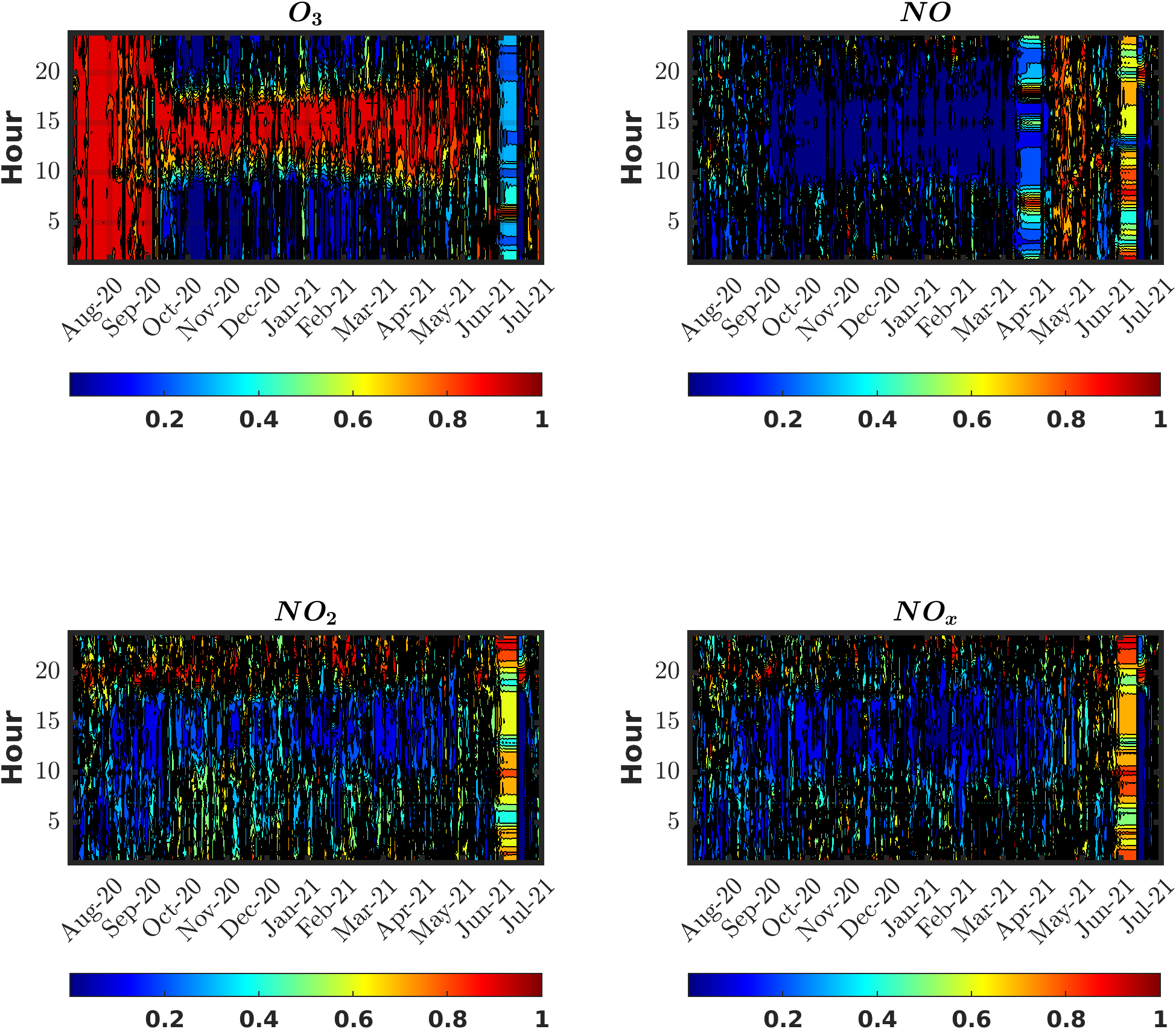}
\caption{Figure shows the time series contour plot of diurnal variation of $O_3$, $NO$, $NO_2$, and $NO_x$ at Industrial for the full time period considered.}\label{fig8}

%The strong diurnal variations in $NO_2$ and PM2.5 are indicative of rush hour related emissions and photochemical production in case of $O_3$.}
\end{figure*}
Here, $x^{(k)}(d,h)$ denote the concentration of $k$-th pollutant at any day $d$ and hour $h$ whereas $x^{(k)}(h)$ is the averaged value of the pollutant's concentration over the year at any specific hour $h$.
Figure~\ref{fig7} and ~\ref{fig8} depicts the diurnal variation as well as monthly variation of $O_3$, $NO$, $NO_2$, and $NO_x$ at Commercial and Industrial, respectively, in the form of a contour plot of dataset from August 2020 to July 2021. Colorbar shows the concentration level normalized to $1$ by the equation~\ref{eqn1}. For $O_3$, one can see the clear dependence of concentration level on the presence of sunlight and hence, on the temperature which is same throughout the year except for monsoon when the concentration level of $O_3$ goes to minimum due to absence of sunlight. Also, within a day, $O_3$ concentration goes to minimum at the nighttime. As already explained in Figure~\ref{fig3} $NO$, $NO_2$, and $NO_x$ shows the opposite nature to $O_3$ for both the measuring stations. $NO$, $NO_2$, and $NO_x$ concentration goes higher during the more traffic rush hours indicating the origin as vehicular and industrial emission. These figures are mainly gives the total information that is shown in Figure~\ref{fig2} and ~\ref{fig3} but with different representation to make the information more clearly visible.
\end{document}